\documentclass[fleqn,usenatbib]{mnras}

\usepackage{mathptmx}

\usepackage[T1]{fontenc}
\usepackage{ae,aecompl}

\usepackage{graphicx} 
\usepackage{amsmath} 

\usepackage{amssymb} 
\usepackage{amsmath}
\usepackage[utf8]{inputenc}

\newcommand*{\subscript}[1]{\ensuremath{_\textrm{{\scriptsize #1}}}}

\title[Puzzling Blue Dips in Swift J1357.2$-$0933]{Puzzling Blue Dips in the Black Hole Candidate \\
Swift J1357.2$-$0933, from ULTRACAM, SALT, ATCA, \textit{Swift} and \textit{NuSTAR}.}

\author[J. A. Paice et al.]{J. A. Paice$^{1,2}$\thanks{E-mail: j.a.paice@soton.ac.uk (KTS)},
	P. Gandhi$^{1}$,
	P. A. Charles$^{1}$\thanks{Leverhulme Emeritus Fellow},
	V. S. Dhillon$^{3,4}$,
	T. R. Marsh$^{5}$, \newauthor
	D. A. H. Buckley$^{6}$,
	M. M. Kotze$^{6,7}$,
	A. Beri$^{1,8}$, 
	D. Altamirano$^{1}$,\newauthor
	M. J. Middleton$^{1}$,
	R. M. Plotkin$^{9,10}$, 
	J. C. A. Miller-Jones$^{9}$, 
	D. M. Russell$^{11}$,\newauthor
	J. Tomsick$^{12}$,
	W. D\'iaz-Merced$^{6}$,
	and R. Misra$^{2}$
	\\
	$^{1}$School of Physics and Astronomy, University of Southampton, Highfield, Southampton, SO17 1BJ, UK\\
	$^{2}$Inter-University Centre for Astronomy and Astrophysics, Pune, Maharashtra 411007, India\\
	$^{3}$Department of Physics and Astronomy, University of Sheffield, Sheffield, S3 7RH, UK\\
	$^{4}$Instituto de Astrofisica de Canarias, E-38205 La Laguna, Tenerife, Spain\\
	$^{5}$Astronomy and Astrophysics Group, Department of Physics, University of Warwick, Gibbet Hill Road, Coventry, CV4 7AL, UK\\
	$^{6}$South African Astronomical Observatory, PO Box 9, Observatory, 7935 Cape Town, South Africa\\
	$^{7}$Southern African Large Telescope, PO Box 9, Observatory, 7935, South Africa\\
	$^{8}$DST-INSPIRE Faculty, IISER Mohali, Knowledge city, Sector 81, Manauli PO, Sahibzada Ajit Singh Nagar, Punjab 140306\\
	$^{9}$International Centre for Radio Astronomy Research, Curtin University, GPO Box U1987, Perth, WA 6845, Australia\\
	$^{10}$Department of Physics, University of Nevada, Reno, Nevada 89557, USA\\
	$^{11}$New York University Abu Dhabi, PO Box 129188, Abu Dhabi, UAE\\
	$^{12}$Space Sciences Laboratory, 7 Gauss Way, University of California, Berkeley, CA 94720, USA\\
}

\date{Accepted XXX. Received YYY; in original form ZZZ}

\pubyear{2018}

\begin{document}
	\label{firstpage}
	\pagerange{\pageref{firstpage}--\pageref{lastpage}}
	\maketitle
	
	\begin{abstract}
        We present rapid, multiwavelength photometry of the low-mass X-ray binary Swift J1357.2-0933 during its 2017 outburst. Using several sets of quasi-simultaneous ULTRACAM/NTT (optical), \textit{NuSTAR} (X-ray), XRT/\textit{Swift} (X-ray), SALT (optical) and ATCA (radio) observations taken during outburst decline, we confirm the frequent optical dipping that has previously been noted both in outburst and in quiescence. We also find: 1) that the dip frequency decreases as the outburst decays, similar to what was seen in the previous outburst, 2) that the dips produce a shape similar to that in binary systems with partial disc occultations, 3) that the source becomes significantly \textit{bluer} during these dips, indicating an unusual geometry compared to other LMXB dippers, and 4) that dip superposition analysis confirms the lack of an X-ray response to the optical dips. These very unusual properties appear to be unique to Swift J1357.2$-$0933, and are likely the result of a high binary inclination, as inferred from features such as its very low outburst X-ray luminosity. From this analysis as well as X-ray/optical timing correlations, we suggest a model with multi-component emission/absorption features with differing colours. This could include the possible presence of a sporadically occulted jet base and a recessed disc. This source still hosts many puzzling features, with consequences for the very faint X-ray transients population. 
		
	\end{abstract}
	
	
	
	
	\section{Introduction}
	
	
	As some of the most extreme environments in the universe, black holes in binary systems can give rise to complex, high-energy sources, and have thus garnered much scrutiny over the past few decades. Low-Mass X-ray Binaries (LMXBs), systems with a black hole being orbited by a companion (or `donor') star, are key in this field; their relatively small size allows them to evolve much more rapidly than their supermassive brethren, and their activity during prolific transient X-ray `outbursts' of enhanced accretion activity is a subject of intense research. Of the handful of systems ($\sim$60, \citealt{corral-santana_blackcat:_2016}) that have been currently identified, some have proven to be more enigmatic than others; Swift J1357.2$-$0933 is a prime example.
	
	Swift J1357.2$-$0933 (hereafter J1357) is an LMXB that was discovered during its 2011 outburst, and was quickly found to have some remarkable properties. Lying at a distance between 2.3 -- 6.3 kpc (\citealt{shahbaz_evidence_2013}, \citealt{sanchez_swift_2015}), it is a Black Hole Candidate (BHC) whose mass is estimated as $\gtrsim$9.3 $M_{\sun}$ \citep{corral-santana_blackcat:_2016}. It is also one of the faintest of the black-hole LMXBs, with a peak luminosity of 1.1${\times}$10$^{35}$ erg s$^{-1}$ at outburst and a very low $L_{X}$/$L_{opt}$ ratio (57) \citep{corral-santana_black_2013}. This puts it into the growing population of Very Faint X-ray Transients (VFXTs), whose peak luminosities are $L_{X}^{peak}$ < $10^{36}$ erg $s^{-1}$ \citep{wijnands_very_2005}; while this has been seen numerous times in neutron star sources, this is a unique property among currently known galactic BHCs.
	
	Swift J1357.2$-$0933 has been observed to go into outburst twice -- once in 2011 (\citealt{krimm_2011}), and again recently in 2017 \citep{drake_2017}. However, even during quiescence, an additional component in the system dominates the light from the companion (\citealt{shahbaz_evidence_2013}, \citealt{russell_optical_2018}). For this reason, the mass of the system has not been dynamically determined; instead, it was inferred by measuring the Full Width at Half-Maximum (FWHM) of the H$\alpha$ profiles from the accretion disc. The companion's properties have been inferred through similar processes and upper limits on its quiescent brightness; it has been found to have a mass of $\sim$0.4 $M_{\sun}$, a binary period of 2.8 $\pm$ 0.3 hours, and spectral type M4.5 (\citealt{corral-santana_black_2013}, hereafter JCS13, \citealt{sanchez_swift_2015}).
	
    Curious, semi-periodic drops in optical flux have been noted in J1357 during both outburst and quiescence (JCS13, \citealt{shahbaz_evidence_2013}). When seen in the 2011 outburst, the frequency of these dips decreased as the outburst declined. JCS13 explain these dips by suggesting that the compact object is surrounded by a thick, irregular torus-like structure, which is seen at a high inclination ($\gtrsim$70$^{\circ}$). Our view of the central emitting region is thus occasionally obscured by outer parts of the accretion disc (\citealt{torres_vlt_2015}, \citealt{sanchez_swift_2015}, \citealt{armas_padilla_x-ray_2014}). The change in frequency could therefore be due to these perturbations travelling outwards from the compact object as the surrounding structure recedes.
	
	However, the exact nature and geometry of the system that creates these dips is still unclear, as is a satisfactory physical explanation. An extended, accretion-disc corona is assumed to be common among X-ray binaries, particularly those hosting a neutron star \citep{white_structure_1985}, yet the prevalence of such structures in black hole systems remains unclear. Alternative interpretations have also been suggested, including a thick inner torus, a warped disc, or an asymmetric outer disc with a tidal arm (JCS13, \citealt{sanchez_swift_2015}).
	
    None of these explanations are fully self-consistent, however, and there are some features which present problems; in particular, the lack of eclipses by the companion star given the inferred high inclination, or the fact that these dips are not present in X-rays (JCS13, \citealt{armas_padilla_x-ray_2014}). Additionally, recent studies report a lack of X-ray spectral features \citep{beri_black_2019}; X-rays arising in the central regions are commonly seen to reflect off physical structures in the accretion environment. While reflection features are strongest in a face-on geometry, and thus at a minima in edge-on geometry, this does not fit the explanation of a toroidal structure where there should be features seen from X-rays passing through the material or reflecting off the back wall (see \citealt{Bauer_NuSTAR_2015}). The material that makes up this obscuring structure is likewise unknown. In short, there are still many aspects of this system that need to be explained.
	
	This paper presents new simultaneous optical, X-ray, and radio data obtained during the decline of the 2017 outburst. We investigate the dips, highlighting their varying colours and response in the X-rays, and compute DCFs between the optical and X-ray data, amongst other work. We thus consider the implications of these results with regards to current models.

	
	\section{Observations}
	
	\begin{table*} 
		\centering
		\caption{Swift J1357.2-0933 Observing log. Epochs refer to the times marked in Figure \ref{fig:timeline}.}
		\begin{tabular}{ccccccc}
			\hline
			Instrument/Telescope & Date & Start (UT) & End (UT) & Start (MJD) & End (MJD) & Epoch \\
			\hline
			\textit{NuSTAR} & 2017-04-28 & 13:09 & 10:13 (+1 day) & 57871.54816 & 57872.42582 &  \\
			XRT/\textit{Swift} & 2017-04-28 & 14:49 & 16:39 & 57871.61721 & 57871.69349 & 1 \\
			SALT & 2017-04-28 & 20:02 & 20:47 & 57871.83477 & 57871.86618 &  \\
			\\
			ATCA & 2017-05-15 & 09:56 & 17:47 & 57888.41388 & 57888.74119 &  \\
			XRT/\textit{Swift} & 2017-05-15 & 12:06 & 12:21 & 57888.50399 & 57888.51454 & 2 \\
			SALT & 2017-05-15 & 18:38 & 19:19 & 57888.77631 & 57888.80485 &  \\
			\\
			SALT & 2017-05-22 & 18:16 & 18:41 & 57895.76112 & 57895.77873 & 3 \\
			\\
			\textit{NuSTAR} & 2017-06-10 & 13:40 & 03:40 (+1 day) & 57914.56944 & 57915.15278 &  \\
			XRT/\textit{Swift} & 2017-06-10 & 14:42 & 15:45 & 57914.61247 & 57914.65659 & 4 \\
			ULTRACAM/NTT & 2017-06-11 & 00:30 & 04:00 & 57915.02083 & 57915.16667 &  \\
			\\
			SALT & 2017-07-19 & 18:27 & 19:22 & 57953.76900 & 57953.80714 & 5 \\
			\hline
		\end{tabular}
		\label{tab:observations}
	\end{table*}
	
	\begin{figure}
		\includegraphics[width=\columnwidth]{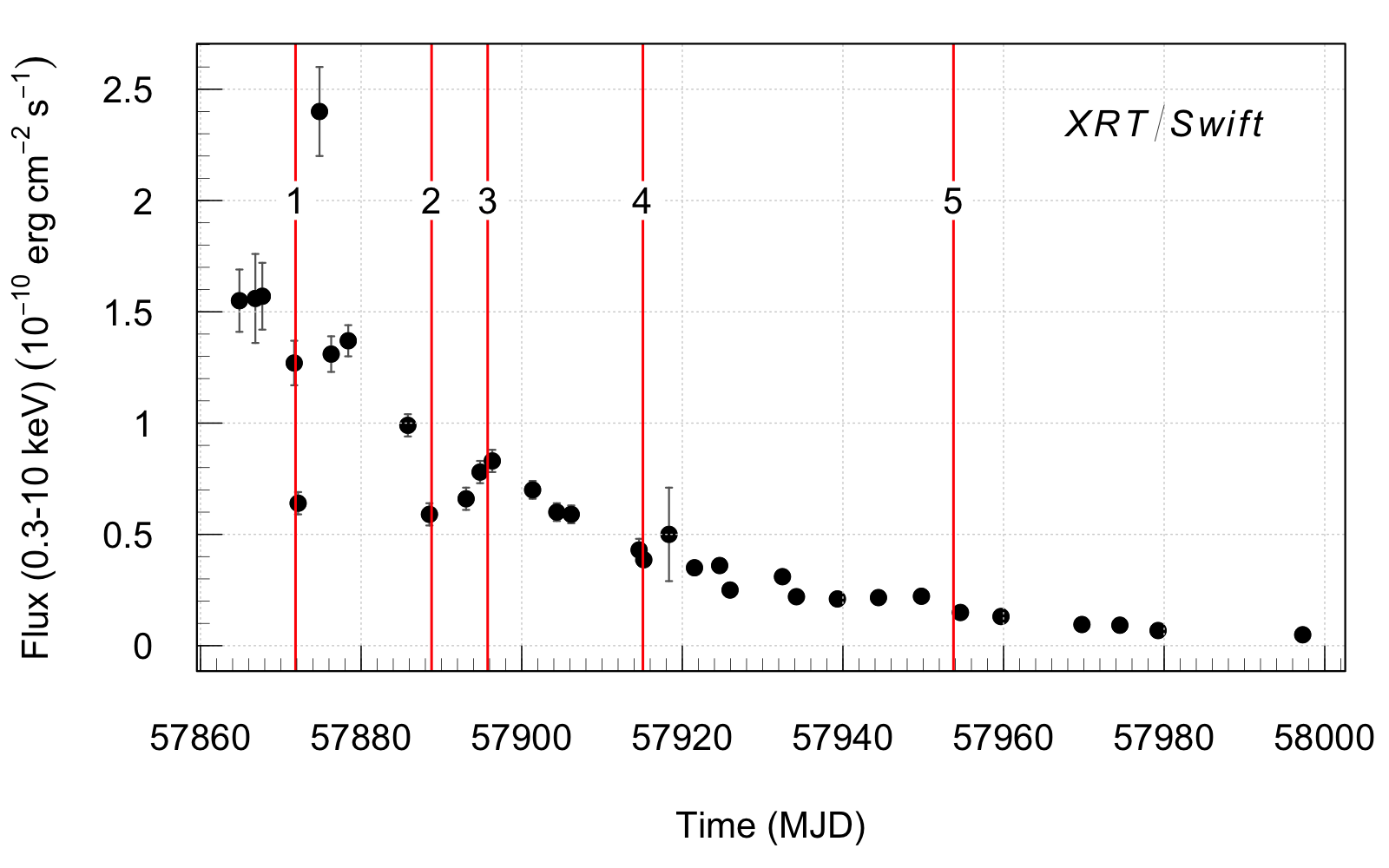}
		\caption{Timeline of Swift J1357.2$-$0933's 2017 outburst, as seen by XRT/\textit{Swift}. Red lines denote dates of observations (see Table \ref{tab:observations}); line 1 is the Apr 28 SALT + \textit{NuSTAR} observation, line 2 is the May 15 SALT + \textit{Swift} + ATCA observation, and line 4 is the Jun 10 ULTRACAM + \textit{NuSTAR} + \textit{Swift} observation.}
		
		\label{fig:timeline}
	\end{figure}

	\subsection{ULTRACAM/NTT -- Fast Optical timing}
	\label{sec:ULTRACAM/NTT} 

	ULTRACAM is a fast-timing optical instrument on the New Technology Telescope (NTT) in La Silla, Chile. It was built for the purpose of fast optical timing in multiple wavebands. To this end, it includes three channels for simultaneous multiwavelength monitoring (with replaceable filters), and it can also observe at frame-rates well above 100Hz - this is achieved by the lack of a physical shutter, and frame-transfer CCDs that can rapidly shift charge into a storage area for reading out, freeing up the original pixels for observation and achieving low dead times \citep{dhillon_ultracam:_2007}.
	
    Observations of J1357 were taken on the night of 2017 June 10, encompassing just over three hours (see Table \ref{tab:observations}). The source was monitored in SDSS $u'$, $g'$ and $r'$ filters, and the times were chosen to coincide with \textit{NuSTAR} observations (see Section \ref{sec:NuSTAR}). ULTRACAM was used in two-window mode (one each for the target and the comparison star), with both window sizes of 50 x 50 pixels with a 2x2 binning for sensitivity and speed. $r'$ and $g'$ bands were observed with an exposure time of 86.1 ms, and a total cycle time of 110.1 ms, giving 24 ms of dead time and a sampling rate of $\sim$9 Hz. J1357 was very faint in $u'$, however, and so ULTRACAM's co-adding feature was used; this combines multiple observations to increase signal-to-noise ratio. For our observation, 20 co-adds were used, giving an exposure time of 2179 ms and thus a frequency of $\sim$0.45 Hz in the blue band.
	
	The data were reduced using the ULTRACAM pipeline v9.14 \citep{dhillon_ultracam:_2007}. The bias was subtracted from each frame, and flat field corrections were also applied. Apertures of optimally varying sizes were used, with radii between 3.15$\arcsec$ and 4.9$\arcsec$, with an annulus of between 6.3$\arcsec$ and 8.75$\arcsec$ to calculate the background, varying with the seeing, which was between 1-2$\arcsec$. These apertures had variable centre positions that tracked the centroids of the sources on each frame, with a two-pass iteration (where an initial pass is made to track the sources on the CCD before a second photometry pass) used for accuracy. Our times were then adjusted to Barycentric Dynamical Time (BJD\_TDB) using methods given in \cite{eastman_achieving_2010}. Interstellar extinction along the line-of-sight to the source in A(V) was found to be 0.094/0.136/0.174 mag in $r'$/$g'$/$u'$ respectively, using \citet{Schlafly_Measuring_2011}; these corrections were applied to our data so as to obtain the intrinsic source magnitudes.
	
	Our comparison star is located at RA = 13:57:18.58, Dec = $-$09:31:20.74 (J2000), and is listed in the Sloan Digital Sky Survey (SDSS) release DR14 as ID 1237671956450377819   \citep{abolfathi_fourteenth_2018} with $r'$/$g'$/$u'$ magnitudes of 13.85/14.50/16.33 respectively. The star was assumed to be constant, and was used for photometric calibration. 
	
	

	\subsection{SALT -- High Speed Photometry}
	\label{sec:SALT} 
	
	The Southern African Large Telescope (SALT) is a 10m-class optical telescope operated by the South African Astronomical Observatory. It was built with spectroscopy in mind, although it can also achieve high-speed photometry (most salient for this paper), and it saw first light in 2005 \citep{buckley_completion_2006}.
	
	High-speed photometry of \textit{Swift} J1357.2$-$0933 was undertaken using the Robert Stobie Spectrograph (RSS; \citealt{Kobulnicky2003}) on 2017 April 28, May 15, 22 and July 19 with seeing 1.5$\arcsec$, 1.8$\arcsec$, 1.3$\arcsec$ and 1.7$\arcsec$ respectively (see Table \ref{tab:observations}). Fast imaging observations were performed in ``slotmode", with a clear, fused silica filter and employing 6 $\times$6 pixel binning, with 100 ms time resolution. This mode is similar to slotmode implemented on the imaging camera, SALTICAM  (e.g. see \citet{O'Donoghue2006}). For RSS slotmode, an occulting mask with a narrow slot is placed at the focus of the telescope, which is then reimaged by the RSS optics onto the mosaic of three edge-butted frame transfer CCDs (E2V 42-81). The slot image has a width of 144 unbinned 15$\mu$m pixels, approximately 20 arcsec on sky, and a length of 8 arcmin. At the end of each slotmode exposure the image is rapidly (in a few ms) moved across the frame transfer boundary of the CCD, and a new exposure is initiated. No shutter is used in slotmode. The images eventually migrate down to the serial readout register in a stepwise manner after each exposure and read out during an exposure. The start times of two consecutive exposures differ by 104 ms, which is the effective time resolution of our observations.
	
	These slotmode observations allowed for the placement of both the target star and an appropriate nearby (1.8 arcmin) comparison star, SDSS J135716.43$-$093140.1 (RA = 13:57:16.452, Dec = $-$09:31:40.14, $g'$ = 15.8), within the slot by rotating its long axis to a position angle of 85$^{\circ}$ using the instrument rotator on SALT. Both stars were imaged onto the central CCD of the detector and so readout by the same CCD amplifier, reducing instrumental effects, such as gain or bias variations. The data were reduced using the PySALT pipeline \citep{Crawford2010}, which corrects for bias, overscan, crosstalk and gain, and differential aperture photometry was undertaken using standard IRAF tasks.
	
	In determining the source magnitude during this observation, the clear filter was approximated as $g'$ by using the comparison star's $g'$ magnitude as a reference.  We estimate that this will produce a systematic uncertainty of $\pm$10\%.

	\subsection{\textit{Swift} -- X-ray}
	\label{sec:Swift}
	
	The XRT onboard the \textit{Neil Gehrels Swift Observatory} \citep{Gehrels_Swift_2004}, was also used to measure the soft X-ray flux of the source (1 -- 10 keV, \citealt{burrows_xrt_2005}). We used three observations with Obs IDs 00088094002 (Apr 28), 00031918058 (May 15), and 00031918066 (Jun 10), coincident with other observations (see Table \ref{tab:observations}). The April 28 observation was made in photon counting mode, while the latter two were made in windowed timing mode.
	
	The April 28 data were processed using the Build XRT Products tool \citep{Evans_Swift_2009}. The other two datasets were processed using {\sc xrtpipeline}, with source and background spectra extracted using {\sc xselect}. A circular region of radius 35$\arcsec$ was used to extract the source, and a similar region centered on an area of no source events was then extracted as the background. Ancillary response files were then created using {\sc xrtmkarf}. The source was found to have an average count rate in the 1 -- 10 keV band of 1.11 $\pm$ 0.03 counts s$^{-1}$ for the Apr 28 observation, 0.81 $\pm$ 0.03 counts s$^{-1}$ for May 15, and 0.67 $\pm$ 0.03 counts s$^{-1}$ for Jun 10.

	\subsection{\textit{NuSTAR} -- X-ray}
	\label{sec:NuSTAR}
	
	The Nuclear Spectroscopic Telescope ARray (\textit{NuSTAR}) is a NASA X-ray satellite that was launched in 2012. Among X-ray telescopes, NuSTAR is particularly notable for its deployable mast, providing a long focal length of 10m when extended; this allows it to focus high-energy X-rays in the 3 -- 78 keV range, the first orbital mission to achieve this \citep{Harrison_NuSTAR_2013}. \textit{NuSTAR} carries two telescopes, FPMA \& FPMB. Except for subtle differences in their effective areas, the two modules are very similar, and for all but spectral analysis we sum the counts from the two here.
	
	Two observations carried out by \textit{NuSTAR} are used here; one on April 28 (Obs ID 90201057002, coincident with \textit{Swift} and SALT) and one on June 10 (Obs ID 90301005002, coincident with \textit{Swift} and ULTRACAM) (see Table \ref{tab:observations}). Due to the low-Earth orbit of \textit{NuSTAR}, the observation was interrupted by frequent Earth occultation and by passages through the South Atlantic Anomaly (SAA), splitting the observation into numerous discrete segments.
	
	Data reduction was completed using {\sc nupipeline}, and source and background regions were selected with a radius of 30$\arcsec$, as recommended by the pipeline documentation. With {\sc nuproducts}, both source and background lightcurves were extracted from both FPMA and FPMB. The background was subtracted manually, and the lightcurves were adjusted to BJD\_TDB using the {\sc ftools} command {\sc barycorr}.

	\subsection{ATCA and EVN -- Radio}
	\label{sec:Radio}
	
    Radio observations of this outburst were carried out with the Australia Telescope Compact Array (ATCA), under project code CX385 (PI Plotkin). We observed on 2017 May 15 from 09:30--18:00 UT, to be coincident with \textit{Swift} and SALT.  Observations were carried out simultaneously in two frequency bands, centred at 5.5 and 9.0\,GHz, each with 2048\,MHz of bandwidth. The array was in the extended 6A configuration.  The bright extragalactic calibrator source PKS 1934$-$638 was used as a bandpass calibrator and to set the flux density scale, and PKS 1406$-$076 to set the time-dependent complex gains. Data processing was carried out according to standard procedures within the Common Astronomy Software Application \citep[CASA v5.1.1;][]{McMullin2007}. The data were imaged using the {\sc CASA} task {\sc clean}, using two Taylor terms to model the frequency dependence over the large fractional bandwidth. Imaging was performed using Briggs weighting with robust=1, to reduce sidelobes from other sources within the field. Flux densities were then measured using {\sc imfit}, requiring a point source during the fitting.

    Following the ATCA radio detection, we requested high angular resolution Very Long Baseline Interferometry (VLBI) observations via a Target of Opportunity proposal on the European VLBI Network (EVN). The project (code RM010; PI Miller-Jones) was observed on 2017 May 21, from 17:25--23:30 UT.  Observations were taken at a central frequency of 4.95\,GHz, with a bandwidth of 256\,MHz.  Ten telescopes (Effelsberg, Jodrell Bank Mk II, Westerbork, Medicina, Noto, Onsala 25m, Torun, Yebes Hartebeesthoek, and Shanghai) participated in the experiment.  Amplitude and bandpass calibration was performed using the EVN pipeline, and fringe-finding and hybrid-mapping of the phase reference source was carried out manually using the Astronomical Image Processing System \citep[AIPS, 31DEC17 version;][]{Greisen2003}.  The calibrated data were imaged using natural weighting, and Swift J1357.2$-$0933 was detected at the $5.2\sigma$ level, at $122\pm23$\,$\mu$Jy\,beam$^{-1}$, at a position consistent (within uncertainties) with the reported {\it Gaia} position \citep{Gandhi_Gaia_2019,Gaia2016,Gaia2018}.

	
	\section{Results}
	
	\subsection{Lightcurves}
	
	\begin{figure*}
		\includegraphics[width=\textwidth]{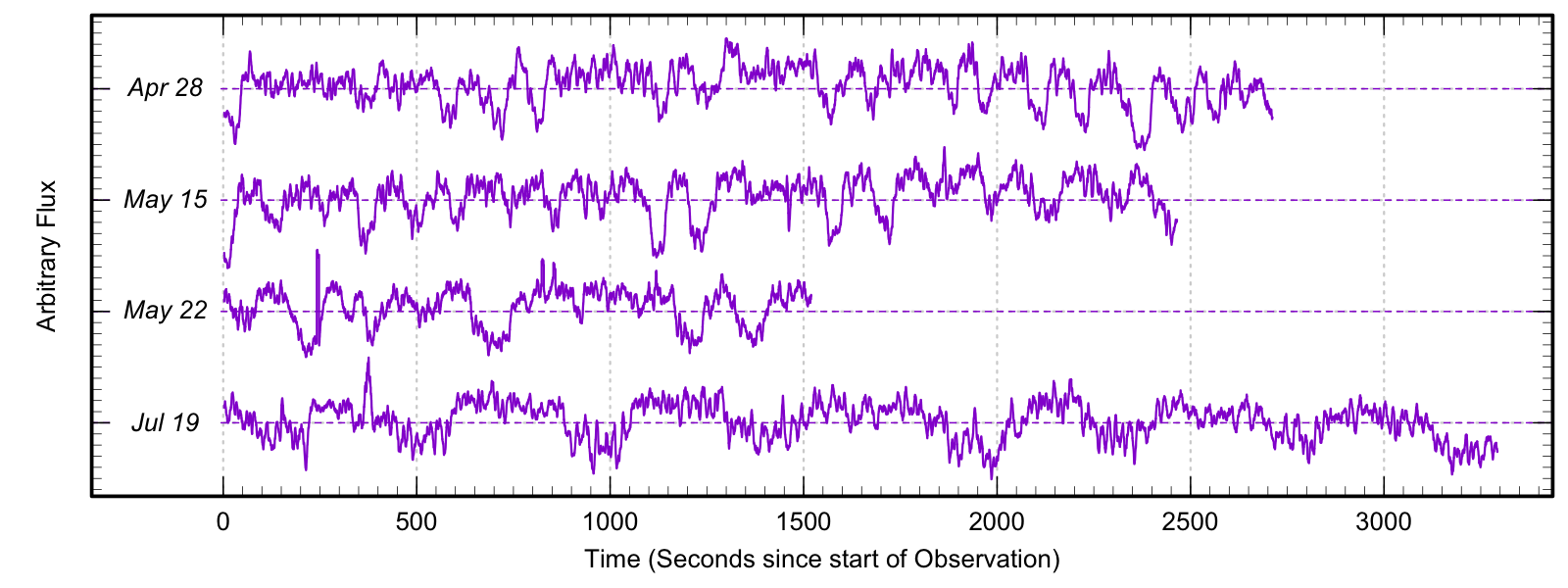}
		\caption{SALT/RSS lightcurves from four different dates. These observations have been binned with a 50s moving average function. Dashed lines show the mean of each observation. Note the dips, and their changing frequency and duration over time.}
		\label{fig:salt_all}
	\end{figure*}
	
	\begin{figure*} \label{sec:lightcurves}
		\includegraphics[width=\textwidth]{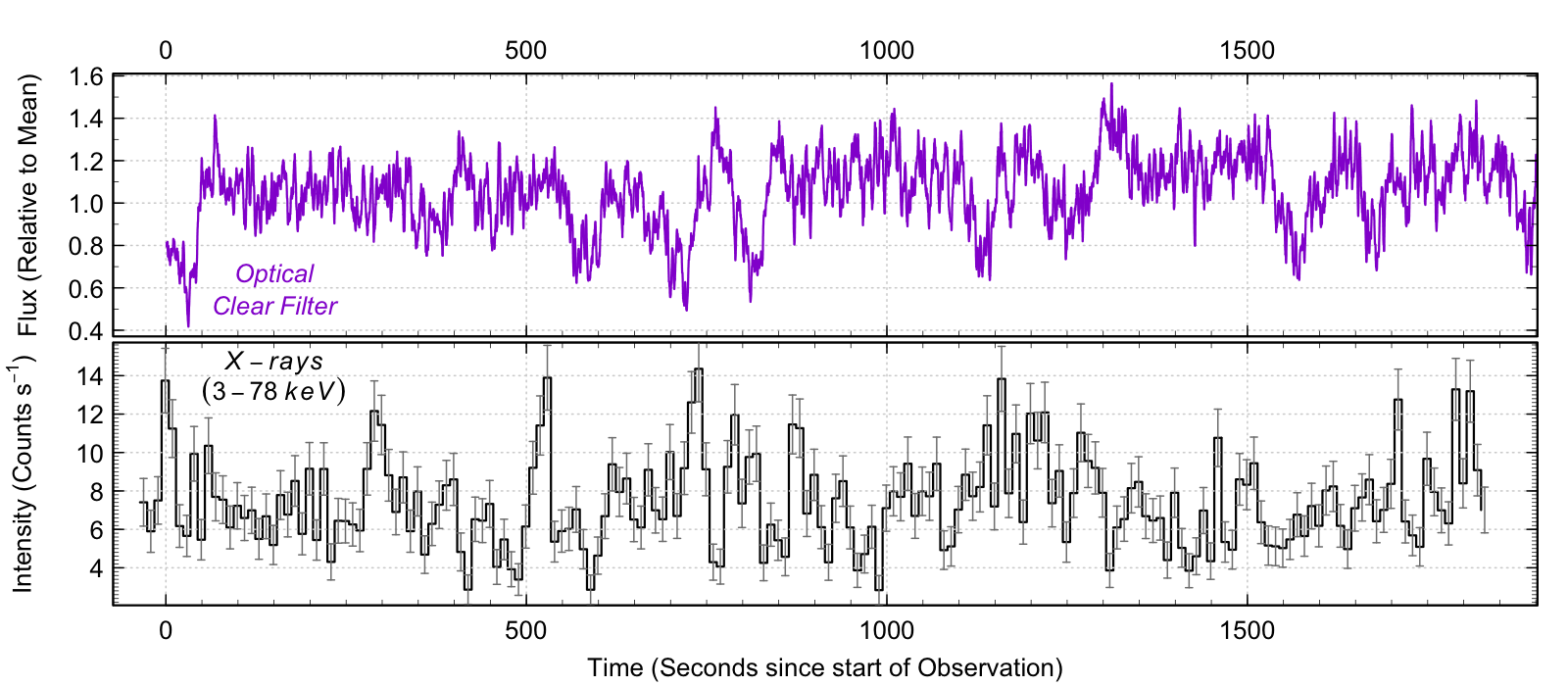}
		\caption{SALT/RSS $\&$ \textit{NuSTAR} lightcurves from 2017 Apr 28.
			\textbf{Top:} Optical from SALT, using a clear filter. These observations have been binned with a 20s moving average function in order to highlight optical dipping.
			\textbf{Bottom:} X-rays from \textit{NuSTAR}, binned every 10s. Errors are shown in grey.}
		\label{fig:salt_nu}
	\end{figure*}
	
	\begin{figure*}
		\includegraphics[width=\textwidth]{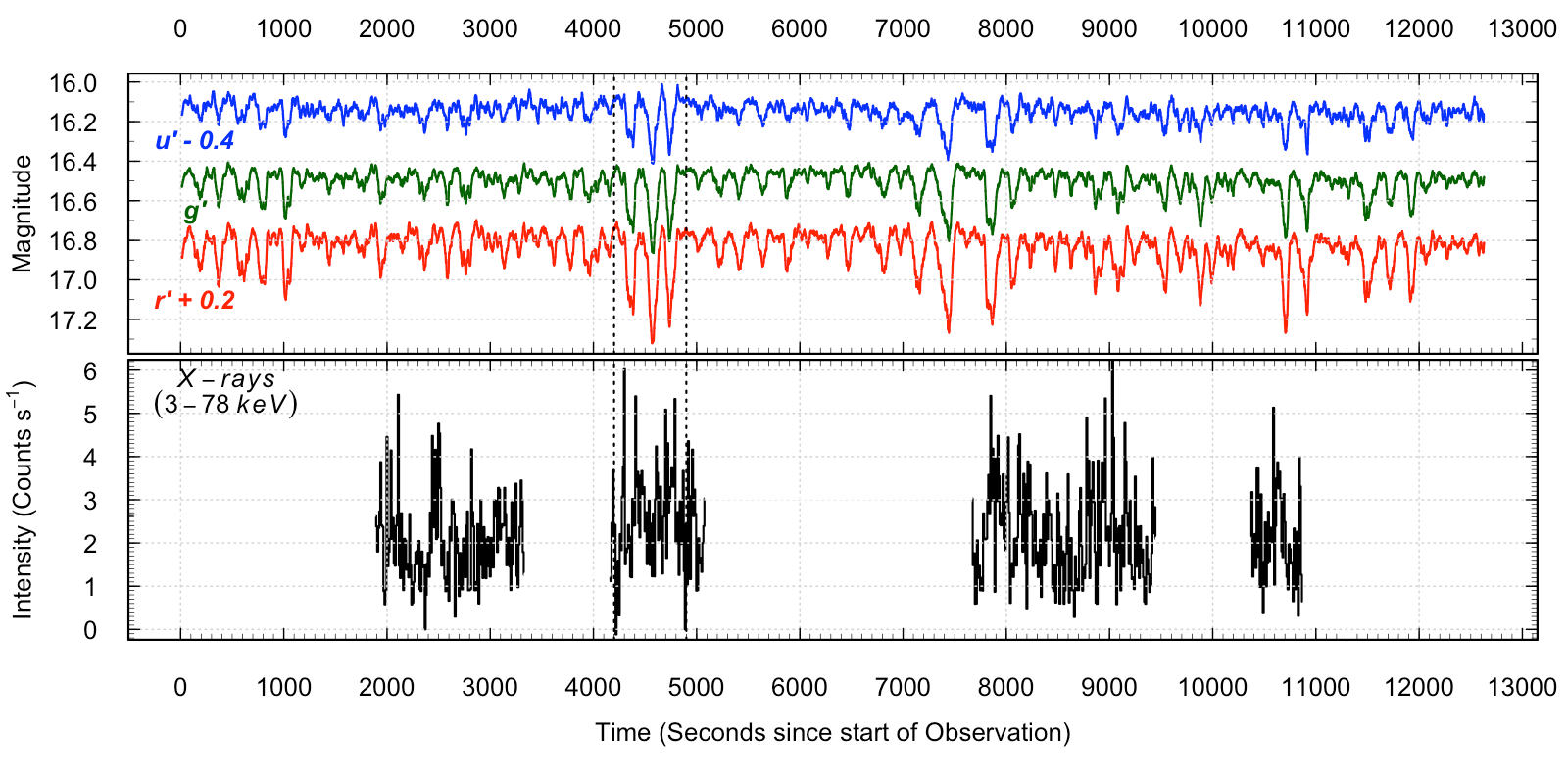}
		\caption{ULTRACAM $\&$ \textit{NuSTAR} lightcurves from 2017 June 10/11.
			\textbf{Top:} Optical: $u'$ (blue, top), $g'$ (green, middle), $r'$ (red, bottom), with $u'$ and $r'$ offset by -0.4 and +0.2 mags respectively for clarity.  A 20s moving average has been applied so as to highlight optical dipping. \textbf{Bottom:} X-ray band, binned every 10s. The dotted lines demarcate the section showing prominent dips, which is expanded in Figure \ref{fig:lc_sec}.}
		\label{fig:lc_full}
	\end{figure*}
	
	\begin{figure*} \label{sec:lightcurves}
		\includegraphics[width=\textwidth]{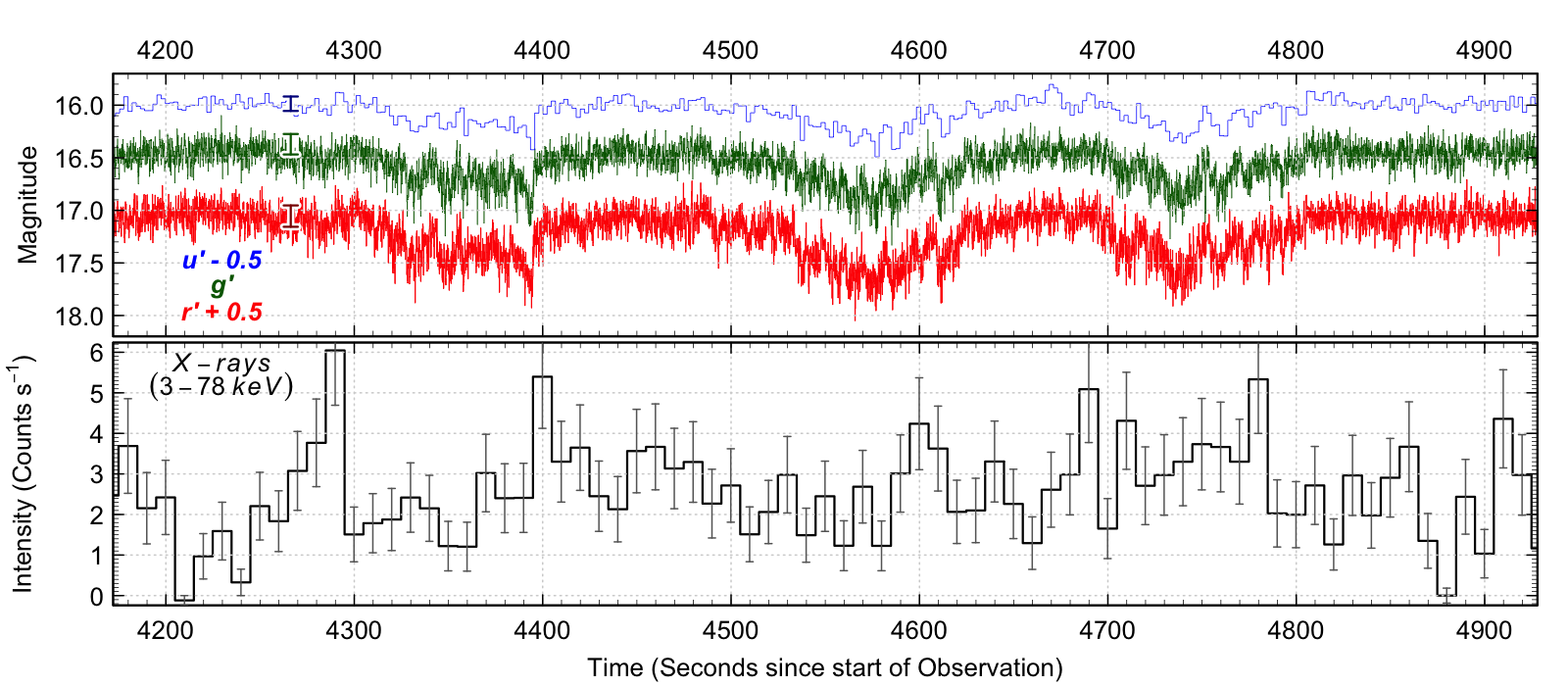}
		\caption{Unbinned section of the Swift J1357.2-0933 lightcurve showing three prominent dips. \textbf{Top:} Optical: $u'$ (blue, top), $g'$ (green, middle), $r'$ (red, bottom), with $u'$ and $r'$ offset by -0.5 and +0.5 mags respectively for clarity. Representative error bars are shown, and note the difference in sampling between the bands. \textbf{Bottom:} X-rays from \textit{NuSTAR}, binned every 10s.}
		\label{fig:lc_sec}
	\end{figure*}
	
	The complete SALT/RSS lightcurves from all four observations are shown in Figure \ref{fig:salt_all}. Dips were present in all four observations, and can be seen to increase in duration and period as the outburst declines. On 2017 April 28/29, a \textit{NuSTAR} observation was coincident with SALT, and had an observed average flux of $\sim$7.4 counts/second in FPMA and FPMB combined; both SALT and \textit{NuSTAR} lightcurves are shown in Figure \ref{fig:salt_nu}, with the X-rays binned every 10 seconds, resulting in $\sim$74 counts per bin.
	
	The complete ULTRACAM and \textit{NuSTAR} lightcurves from 2017 June 10/11 are shown in Figure \ref{fig:lc_full}. Over the course of the night, J1357 periodically varied by 0.3-0.4 mag in all three bands, and the source S/N was >20 per frame in $r'$ for most of the night. Of the 3 bands, the source was brightest in $g'$ ($\sim$16.64) and dimmest in $r'$ and $u'$ ($\sim$16.73 and $\sim$16.72 respectively). Four discrete segments of the \textit{NuSTAR} observation were coincident with ULTRACAM. J1357 was fainter in X-rays compared to the earlier observation, with an observed average flux of only $\sim$2.2 counts/second in FPMA and FPMB combined. The X-rays were again binned every 10 seconds, resulting in $\sim$22 counts per bin.
	
	All optical lightcurves periodically contain large dips, and ULTRACAM lightcurves show them to be simultaneous across all three bands. These dips are not immediately visible in X-rays; this can be seen in detail in Figures \ref{fig:salt_nu} and \ref{fig:lc_sec}. Such dips change in duration, from $\sim$70 seconds on April 28, to $\sim$250 seconds on July 19. ULTRACAM data shows them to drop between 0.1 -- 0.5 magnitude, with a dependence on wavelength. The colour of these dips, and the X-ray response to them, are discussed in Section \ref{dips}.

	
	\subsection{Power Spectral Densities} \label{psd}
	
	\begin{figure}
		\includegraphics[width=\columnwidth]{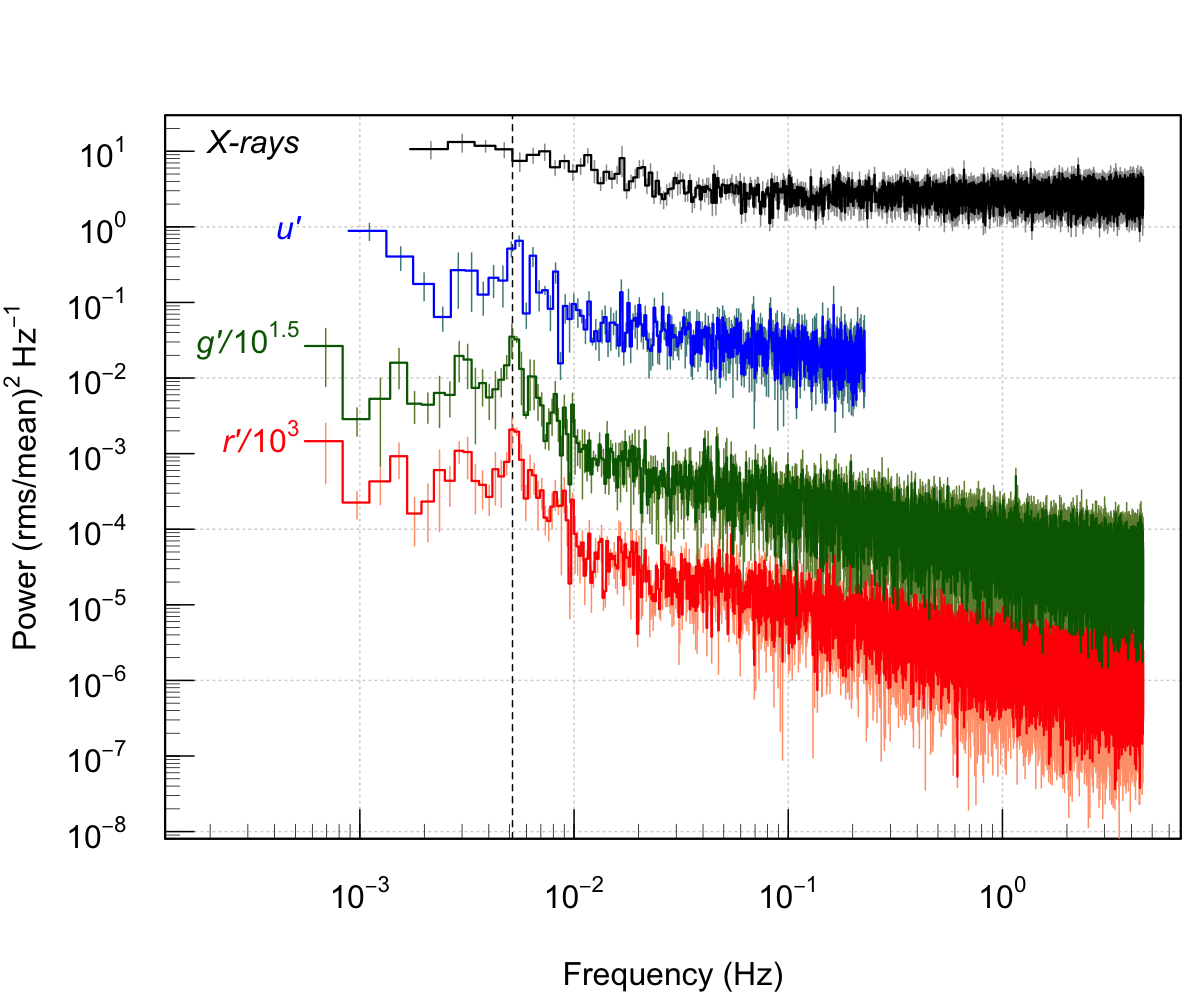}
		\caption{Optical and X-ray PSDs of J1357 with ULTRACAM and \textit{NuSTAR} respectively, showing from top-to-bottom the X-ray (black), $u'$ (blue), $g'$ (green), and $r'$ (red) bands. For clarity, $g'$ has been shifted downwards by a factor of 10$^{1.5}$, and $r'$ by a factor of 10$^{3}$. They were created by splitting the data into fifteen segments of roughly 1165 seconds each in X-rays, five segments of roughly 2250 seconds each in $u'$, and three segments of roughly 3600 seconds each in $g'$ and $r'$. Fourier transforms of each of those segments were then taken, and the results averaged. A black dashed line shows the lorentzian midpoint noted in Table \ref{tab:lor1}.}
		\label{fig:comb_psd}
	\end{figure}
	
	\begin{figure}
		\includegraphics[width=\columnwidth]{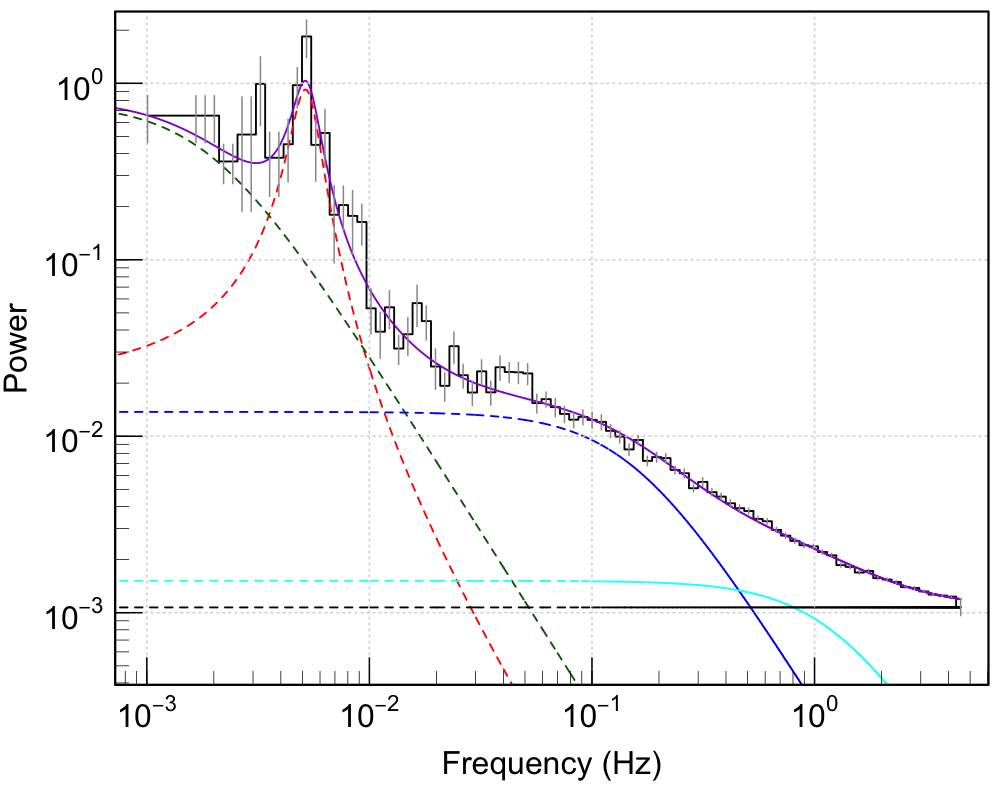}
		\caption{The optical PSD in $r'$, binned logarithmically (black solid line). The purple solid line is a model, made up of four Lorentzians (red, green, blue, and cyan dashed lines) and a constant noise component (black dashed line). Note the QPO at around 5 $\times$ 10$^{-3}$ Hz.}
		\label{fig:optical_psd_lorentzian}
	\end{figure}

	The power spectral densities (PSDs) of the source from \textit{NuSTAR} and ULTRACAM were computed by splitting the data into segments of equal length, applying a fast Fourier transform to the lightcurve of the target, and then averaging the results (see Figure \ref{fig:comb_psd}). The units were rms-normalised $P_{\nu}$ (Power) using the following formula:
	
	\begin{equation}
		\centering
		P_{\nu} = \frac{2 \times E \times N}{ \overline{x}^2}
	\end{equation}
	
	where $E$ is the exposure length per frame (time resolution for X-rays), $N$ is the number of bins in each segment, and $x$ is the counts. Standard errors on the mean were calculated for each bin. For this analysis, the X-ray event data were binned to the $r'$ and $g'$ time bins.
	
	
	The fractional RMS ($F_{rms}$) of the bands was calculated by splitting the lightcurves into ten segments, and then using the following formula:
	
	\begin{equation}
		\centering
		F_{rms} = \sqrt{ \frac{Var(x) - \overline{x_{err}^2}}{ \overline{x}^2}} 
	\end{equation}
	
	where x is the counts, and $x_{err}$ is the error on the counts. The total fractional RMS is defined as the mean of the results, while the error is the standard deviation. The fractional RMS ($F_{rms}$) of the bands are X-ray = 0.409 $\pm$ 0.188, $u'$ =  0.082 $\pm$ 0.014, $g'$ = 0.100 $\pm$ 0.012, and $r'$ = 0.120 $\pm$ 0.018. 
	
	The most striking feature of these figures is the peak in all three optical bands at around 5 $\times$ 10$^{-3}$ Hz, which relates to the approximate frequency of the dips. The X-ray PSD shows no such peak.
	
	To investigate the PSDs further, we used a model composed of Lorentzians (as defined in \citealt{Nowak_Luminosity_1999}), whose form is given by:
	
	\begin{equation}
		\centering
		L(x) = \frac{N}{\pi} \frac{\frac{1}{2}\Gamma}{(x-x_0)^2+(\frac{1}{2}\Gamma)^2} 
	\end{equation}
	
	\noindent where $N$ is the normalisation, $\Gamma$ is the Full Width at Half Maximum (FWHM), and $x_0$ is the midpoint ($x_0$ = 0 indicates a zero-centred Lorentzian).
	
	The optical data were well fitted with four Lorentzians plus a white noise component; this included a strong signal seen at 5.16 $\times$ 10$^{-3}$ Hz, with a Q-factor of 3.25. These are given in Figure \ref{fig:optical_psd_lorentzian}, while the parameters can be seen in Table \ref{tab:lorentz}. For the X-ray data, \citet{beri_black_2019} fitted a single Lorentzian at 4 ($\pm$ 1) $\times$ 10$^{-3}$ Hz (though at a low Q-factor of 0.57). 
	
	\begin{table} 
		\centering
		\caption{Parameters for the four Lorentzians used in Figure \ref{fig:optical_psd_lorentzian}.}
		\label{tab:lor1}
		\begin{tabular}{cccc}
			\hline
			Colour & $N$ ($\times$10$^{-3}$) (Hz$^{-1}$) & $\Gamma$ ($\times$10$^{-3}$) (Hz) & \textit{x$_0$} ($\times$10$^{-3}$) (Hz) \\
			\hline
			Red & 2.30 $\pm$ 0.82 & 1.59 $\pm$ 0.43 & 5.16 $\pm$ 0.37 \\
			Green & 4.71 $\pm$ 1.49 & 3.87 $\pm$ 2.23 & 0\\
			Blue & 6.47 $\pm$ 1.82 & 300 $\pm$ 153 & 0\\
			Cyan & 5.95 $\pm$ 0.52 & 2500 $\pm$ 2049 & 0\\
			\hline
		\end{tabular}
		\label{tab:lorentz}
	\end{table}

	\subsection{Discrete Correlation Functions} \label{ccf}
	
	\begin{figure}
		\includegraphics[width=\columnwidth]{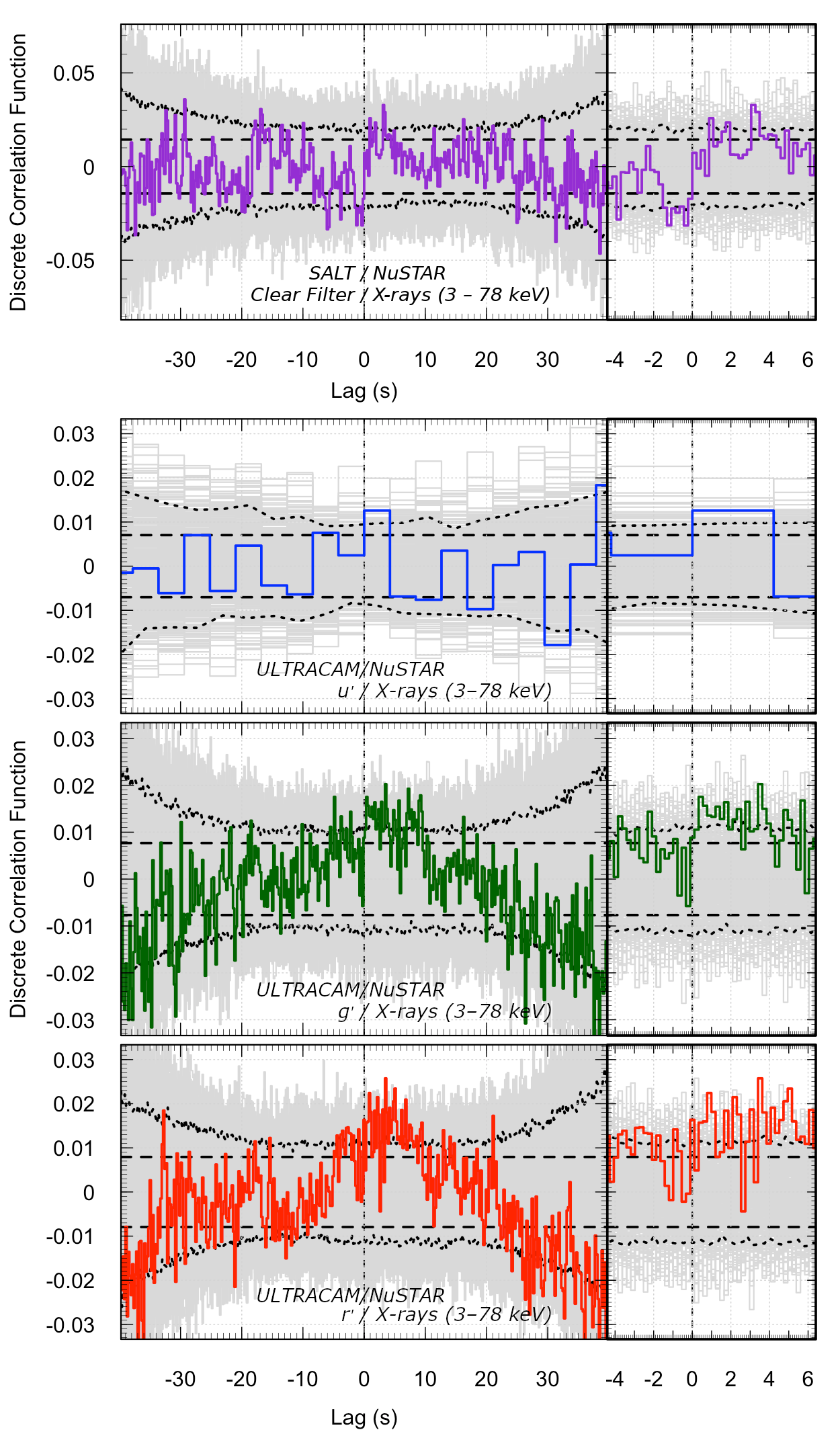}
		\caption{Discrete correlation functions, created from 50s segments. These graphs show the Apr 28 observation (SALT/\textit{NuSTAR}, purple, very top), as well as the Jun 10 observation (ULTRACAM/\textit{NuSTAR}) in all three ULTRACAM bands ($u'$: blue, top, $g'$: green, middle, and $r'$: red, bottom). All plots are optical vs X-ray, i.e., a peak at positive lags means optical features lag X-ray. DCFs created from 500 simulated lightcurves are shown in grey, with their standard deviation over the shown range shown as the black dashed line, and the contours containing 5\%-95\% of the simulations are shown as black dotted lines. A zoom-in about 0s lag is shown on the right for clarity. Note the lack of any clear features, aside from the potential peak directly after 0s lag.}
		\label{fig:dcf50}
	\end{figure}
	
	
	The simultaneous multiwavelength nature of the observations allowed us to use a Discrete Correlation Function (DCF) analysis, which investigates any correlations between different bands. Our analysis used methods presented in \citet{Edelson_DCF_1988}. For this analysis, as with the Power Spectra, the X-ray event data were binned to the $r'$ and $g'$ time bins.
	
	To create a DCF, we split the data into segments of equal size (the size of these segments was varied in order to probe different length scales). After pre-whitening the data to remove any red noise trend \citep{Welsh_Reliability_1999}, we used lag bins with sizes approximately equal to twice the optical time resolution. The final DCF corresponds to the median of all segments. The DCFs were calculated with optical against X-ray signals; hence, a correlation at positive time lags would indicate the optical lagging the X-ray signal.
	
	To analyse the significance of our results, we simulated lightcurves based on our optical data. To do this, we Fourier transformed the lightcurves, randomised the phases (i.e. the arguments of the resulting complex numbers), and then inverse Fourier transformed the result. This simulated lightcurve therefore had the same power spectrum as the source lightcurve, but was randomised in time and would thus be uncorrelated with respect to X-rays. We then found the standard deviations of these simulations, as well as their 5-95\% intervals.
	
	The resultant DCFs are displayed in Figure \ref{fig:dcf50} from the nights of both April 28 and June 10. They show a lack of strongly significant features; however, a peak can be seen in all three bands in the bins directly after 0s. This peak, at its height, lies above the 95\% confidence line (dotted). These results will be discussed further in Section \ref{discussnew}.

	\subsection{Optical Dips -- Superposition} \label{dips}
	
	The frequent dips seen by ULTRACAM, shown in Figure \ref{fig:lc_full}, occur over timescales of around 100 -- 200 seconds from start to finish. During this time the brightness of the source can drop by 0.1 -- 0.5 magnitude, and minima of the dips can occur roughly every 200 seconds (c.f. Section \ref{psd}). However, neither the shape nor the depth of these dips appear constant.
	
	To build up a better picture of this feature, we superimposed a number of dips from the June 10 observation to create an 'average' dip event. Using the $r'$ lightcurves, we first selected every dip that involved a drop of $>$10\% in flux from the median. We then determined the approximate midpoint of each of these events by binning the lightcurve with a moving average function every 200 points. Then, we found the time when this binned lightcurve last dropped below the median value (of the entire lightcurve), and the time when it next went above that same median value. The midpoint of these two times was taken to be the midpoint of the dip.
	
\begin{figure}
		\includegraphics[width=\columnwidth]{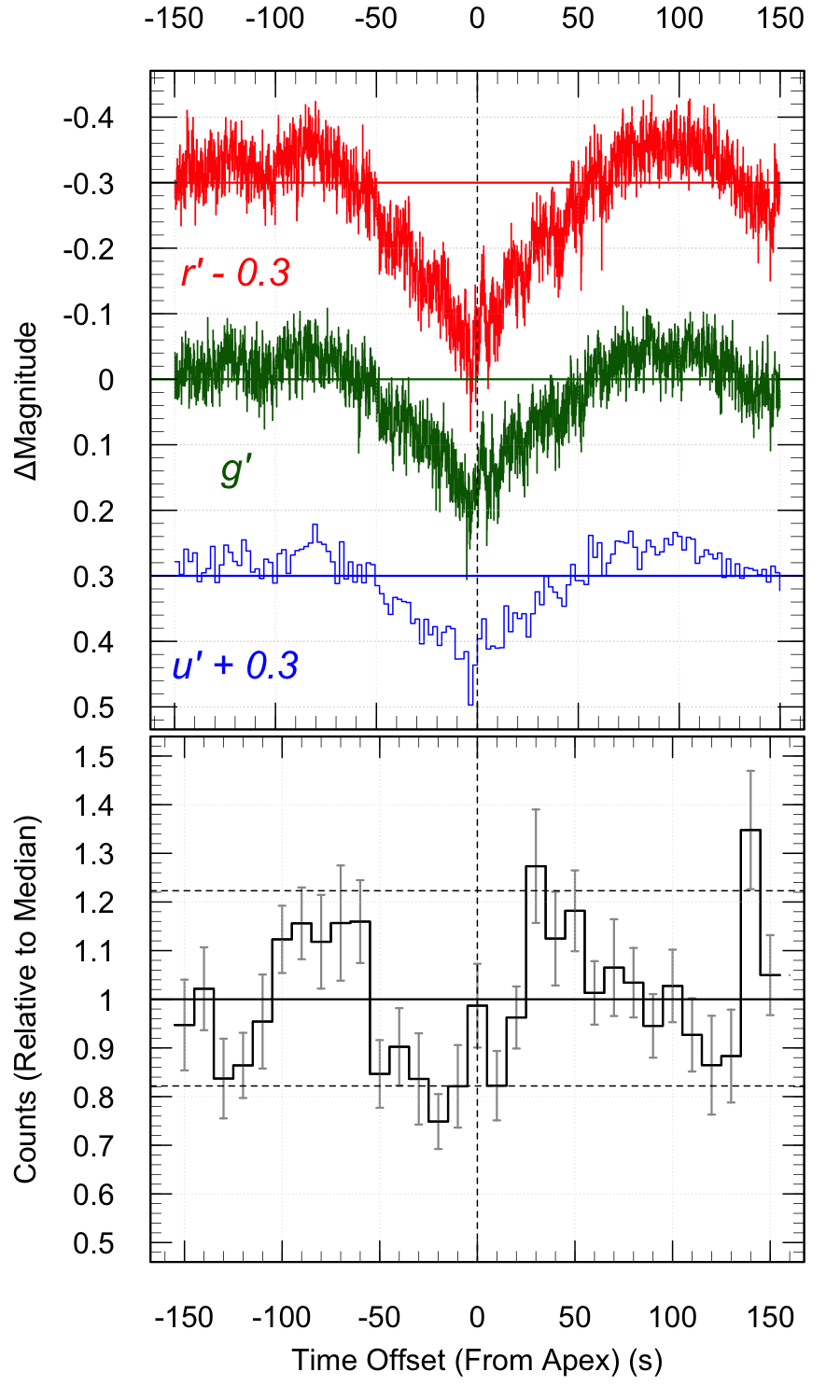}
		\caption{Lightcurve superposition of twelve optical dips that also have X-ray observations, normalised to each band's respective median (shown as a solid line in that band's colour). \textbf{Top:} Optical bands, in the top-to-bottom order $r'$, $g'$, $u'$, with the $u'$ offset by +0.3 mag, and the $r'$ offset by -0.3 mag. Note the smooth shape of the dip, and the increasing depth from $u'$ to $g'$ to $r'$. \textbf{Bottom:} X-ray band, where no significant variations are noted. Dashed lines show the 5-95\% significance intervals of data.}
		\label{fig:super}
	\end{figure}

	Dips were constrained so that only those that had correlated X-ray observations across a 300s range were selected, resulting in twelve dips being selected. These were then averaged, and the result is shown in Figure \ref{fig:super}.
	
	The figure clearly shows dips in all three optical bands, with the red band dropping significantly more than the green, which drops more than the blue. Quantitatively, and on average, the dips relative to the median level are 0.2-0.3 mag ($r'$), 0.15-0.2 mag ($g'$) and 0.1-0.15 mag ($u'$). The average dip's total duration is between 150-180s. It is also implied from Figure \ref{fig:super} that there are no X-ray dips associated with the optical in the 3 -- 78 keV range - the highest deviation is $\sim$0.35 counts s$^{-1}$ from median, and there is an apparent drop in X-ray counts at an offset of -20s, but neither of these well match the shape seen in optical, and are mostly inside the 5-95\% scatter range indicated.
	
	\subsection{Optical Dips -- Evolution} \label{dipevolution}
	
	One of the more remarkable properties of these dips is the evolution of the dip periodicity; JCS13 found that the period increased as the source declined from the outburst peak.
	
	Their interpretation of this dip behaviour was that the inner edge of some obscuring material was moving outwards through the disc during the decline.  Others have since confirmed that these dips exist in quiescence at much lower frequencies (5 $\times$ 10$^{-4}$ Hz, \citealt{shahbaz_evidence_2013}), consistent with the pattern seen during outburst decline.
	
\begin{figure}
		\includegraphics[width=\columnwidth]{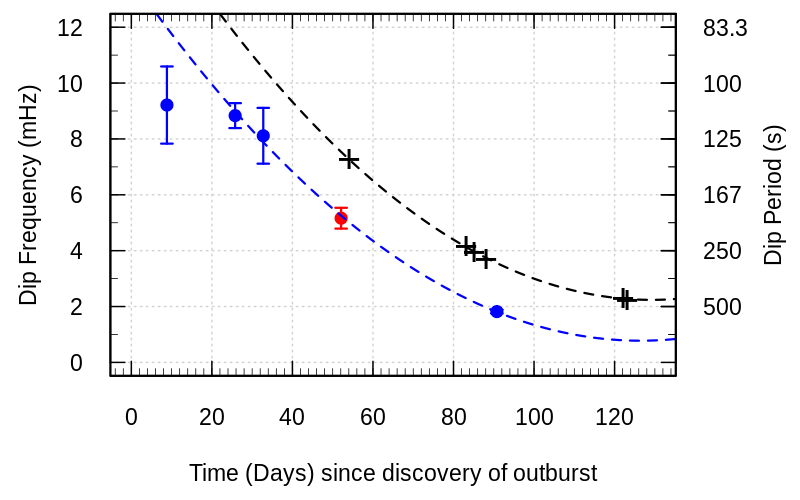}
		\caption{Evolution of the dip frequency over time for both the 2011 and 2017 outbursts. Blue circles mark SALT observations, while the red circle marks the ULTRACAM/NTT observation. Black crosses are the 2011 observations reported in JCS13. All points from both outbursts are plotted against time since each outburst was first reported (\citealt{krimm_2011}, \citealt{drake_2017}).
		The black dashed line shows a fit to the JCS13 data (given in the source paper), while the blue dashed line is a fit to the latter four points of the new data; these are discussed further in Section \ref{discussevolution}.}
		\label{fig:dip_evo}
	\end{figure}

	Our extensive coverage of the 2017 outburst with SALT allows us to follow the evolution of the dip over the course of the outburst. Lorentzians were then fitted to the power spectra to obtain a frequency for the dips in each scenario, and the results plotted in Figure \ref{fig:dip_evo}. The resultant graph shows a decline that is very similar to Figure 3(G) in JCS13, and a similar curve can be fitted the new data - albeit one that cannot simply be shifted in time to fit. These dips are discussed further in Section \ref{discussevolution}.
	
	\subsection{Spectral Energy Distribution} \label{spectra}
	
	Three Spectral Energy Distributions (SEDs) of J1357 were created, one on each of the three dates of correlated observation; April 28 (\textit{NuSTAR}, \textit{Swift}, SALT), May 15 (\textit{Swift}, SALT, ATCA) and June 10 (\textit{NuSTAR}, \textit{Swift}, ULTRACAM). The {\sc xspec} software \citep{Arnaud_Xspec_1996} was used to fit the data, which involved the models {\tt phabs} and {\tt TBabs} (accounting for absorption by the interstellar medium, where standard galactic values were assumed for ISM abundances), {\tt powerlaw} (standard power law), and {\tt diskbb} (black body spectrum from a disk). All errors given are quoted at 1-$\sigma$ confidence. 
	
	For April 28, the \textit{Swift} and \textit{NuSTAR} observations were fit with an absorbed power law, with {\it N$_{H}$} = 2.6 $\pm$ 0.3 $\times$ 10$^{21}$ cm$^{-2}$ and photon index ($\Gamma$) = 1.66 $\pm$ 0.01. A small constant of 1.00561 was also applied to FPMB. This gave a flux (2 -- 10 keV) of 1.01 $\times$ 10$^{-10}$ ergs cm$^{-2}$ s$^{-1}$.
	
	For May 15, the \textit{Swift} observation alone was fit with an absorbed power law, with {\it N$_{H}$} = 2.6 $\pm$ 0.7 $\times$ 10$^{21}$ cm$^{-2}$ and photon index ($\Gamma$) frozen at 1.8. The radio observation was also fit with a power law, which was found to have a spectral index $\alpha$ (F$_{\nu}$ $\propto$ $\nu^{\alpha}$) of 0.47 $\pm$ 0.19.
	
	For June 10, we found that the X-ray emission alone can be well described with an absorbed power law, with {\it N$_{H}$} = 3 $\pm$ 0.6 $\times$ 10$^{21}$ atoms cm$^{-2}$ and $\Gamma$ = 1.81 $\pm$ 0.01. This gave us a flux (2 -- 10 keV) of 3.03 $\times$ 10$^{-11}$ ergs cm$^{-2}$ s$^{-1}$. We also fit the multifilter optical data for this observation; this, and its implications for the jet break, are noted down in Section \ref{discusssed}.

    \subsection{Radio Observations and VLBI position}
	\label{sec:position}
    
    Our observations with ATCA show an inverted spectral index (i.e., F$_{\nu}$ increasing with $\nu$). It has been shown before that J1357 lies significantly off the L$_{Radio}$/L$_{X-ray}$ plane that other hard-state X-ray Binaries follow \citep{plotkin_clean_2016}. Using our present data, we find source radio luminosities (5.5 GHz) of 4.3 $\times$ 10$^{27}$ $-$ 3.3 $\times$ 10$^{28}$ erg s$^{-1}$ and X-ray luminosities (1 $-$ 10 keV) of 3.1 $\times$ 10$^{34}$ $-$ 2.3 $\times$ 10$^{35}$ erg s$^{-1}$, for distances of 2.3 $-$ 6.3 kpc. This places the source in a very similar position in the $L_{Radio}$/$L_{X-ray}$ plane during the current (2017) outburst, as the previous (2011) outburst.

    Our EVN observations have allowed the most accurate determination yet of the source's position, phase referenced to the extragalactic calibrator source J1401$-$0916 \citep{Beasley2002}, whose position was assumed to be RA=14:01:05.331831, Dec = $-$09:16:31.57207 (J2000). Relative to this reference position, Swift J1357.2$-$0933 was found to be at RA= $13:57:16.835810\pm0.000028$, Dec = $-09:32:38.80117\pm0.00047$ (J2000); this position is a distance of 7.34 $\times$ 10$^{-3}$ $\arcsec$ from the optical counterpart given in \textit{Gaia} \citep{Gandhi_Gaia_2019}.

	\section{Discussion} \label{discuss}
	
	\subsection{Current Knowledge and Previous Models} \label{discussprevious}
	
	Since its discovery, J1357 has proven to be highly enigmatic, as no other system has demonstrated the extraordinary variable dipping period that evolves during the outburst. It also has properties that seem to be in conflict with ``standard'' LMXB models, which we shall address here.
	

	
	A key parameter with any LMXB interpretation is its orbital inclination. While the optical dips would suggest a high inclination (JCS13), many of the past investigations on this source have commented on the conflicting evidence regarding this. Low absorption due to the disc \citep{torres_vlt_2015}, soft emission apparently from inner regions of the accretion disc \citep{armas_padilla_x-ray_2014}, a lack of reflection features from the back wall of the torus, and the lack of X-ray dips \citep{beri_black_2019} all imply that a high inclination is unlikely for this source; however, the depth of the absorption core in emission lines \citep{torres_vlt_2015}, He I line cores and the inferred high mass function \citep{sanchez_swift_2015}, as well as a lack of X-ray reflection \citep{beri_black_2019} suggest the opposite. Additionally, the source has been shown to be radio-quiet relative to its X-ray emission \citep{plotkin_clean_2016}, a property that could be consistent with high inclinations in black hole transients \citep{Motta_Radio_2018}.
	
	These conflicting indications await an adequate explanation, but overall, all of these works agree that a reasonably high inclination is most likely, though they disagree on the actual value; \citet{sanchez_swift_2015} suggest that $\geq$80$^{\circ}$ is the most plausible solution (based on He I line cores and mass function), while JCS13 and \citet{torres_vlt_2015} constrain it to between 70--80$^{\circ}$ (citing the optical dips and lack of intrinsic absorption respectively). \citet{stiele_nustar_2018} and \citet{beri_black_2019}, meanwhile, studied the X-ray spectrum of the source under two inclination models, 70$^{\circ}$ and 30$^{\circ}$. The former found no strong evidence supporting either inclination from the spectrum alone, while the latter could not find any evidence to straightforwardly support a high inclination, mostly due to the lack of any reflection features; however, at the same time, they could not achieve a better fit to their results with a lower inclination interpretation.
	
	The optical dips remain one of the key points that these investigations cite in support of a high inclination; since there are no models at present suggesting otherwise, our ansatz is similarly an inclination higher than 70$^{\circ}$. The current model, put forward by JCS13, hypothesises that these dips are caused by vertical extensions of a torus-like structure around the compact object that periodically occult the emitting regions. It is this model that we will investigate with our new data.

	\subsection{Optical Dips -- Colour} \label{discussnew}

	By observing J1357 in three optical bands, we are able to probe its variability as a function of wavelength throughout the observation. The dips present in the optical lightcurve are of particular interest in this regard.
	
	These dips have, on average, a V-shaped lightcurve that can best be seen in Figure \ref{fig:super}. This sort of lightcurve is very similar to those seen in grazing binary systems with gradual source obscuration (e.g. as seen in \citealt{howell_kepler_2010}). Indeed, obscuration of an emitting region has previously been postulated as a cause \citep{armas_padilla_multiwavelength_2013}.
	
\begin{figure}
		\includegraphics[width=\columnwidth]{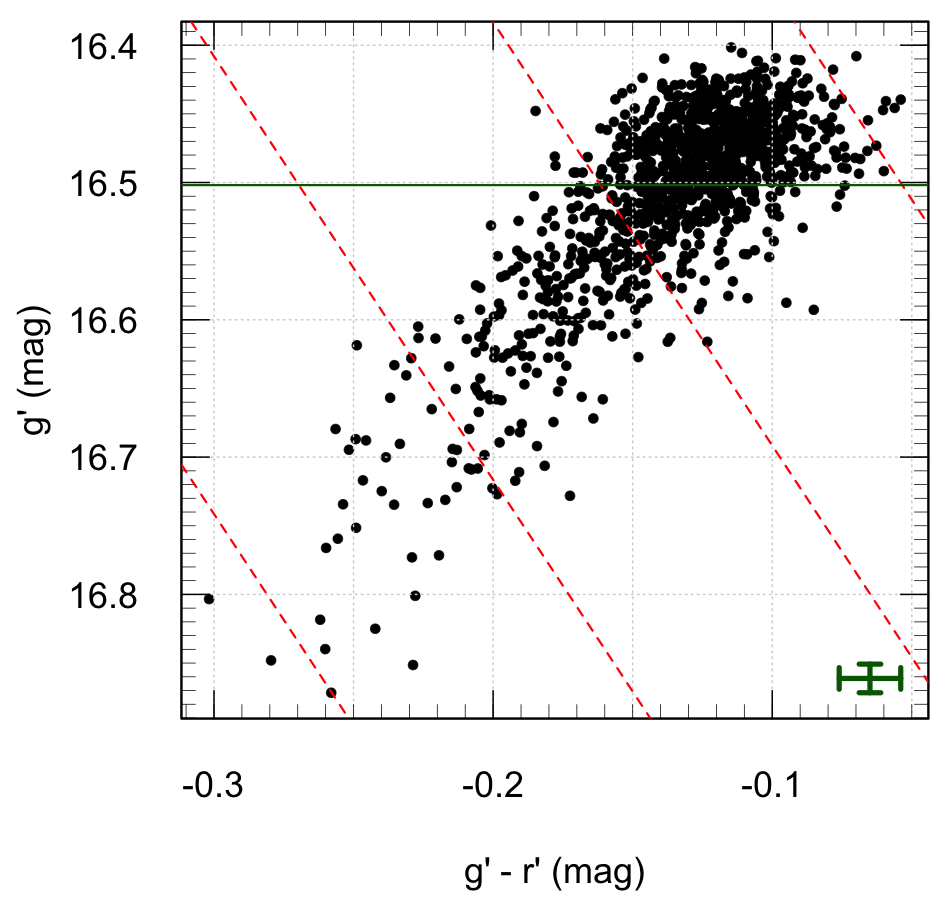}
		\caption{Colour-Magnitude diagram covering the entire optical lightcurve, plotting $g'$ against $g'$ - $r'$. The data have been binned to ~10s resolution so as to minimise scatter and highlight the main trend. The dashed red lines represents the slope that the data should follow under extinction by standard dust, and the horizontal solid line is the $g'$ median magnitude. Representative error bars are shown at the bottom-right. The variations in J1357 are clearly orthogonal to what would be expected through dust extinction.}
		\label{fig:colourt}
	\end{figure}

	What can the colour tell us about this system? As noted earlier in Section \ref{dips}, these dips are significantly blue compared to the rest of the lightcurve. This result is puzzling; standard absorption would affect shorter wavelengths much more than longer, causing a significant reddening, rather than the blue colour seen here \citep{cardelli_relationship_1989}. To illustrate this disparity a colour-magnitude locus was also created, and is shown in Figure \ref{fig:colourt}; this makes it clear that the source becomes bluer, not redder, during the dips. Hence, reddening due to standard absorption cannot be a solution. Moreover, if the obscuring matter were optically thick, we would expect to see achromatic colour changes. Obscuration by the disc would also cause dips in X-rays.

	One possibility is that the observed emission from the source is composed of two components, with each being a different colour. One scenario in which this could occur is if one considers the presence of blue disc and red jet spectral components. Viscous optical disc emission, which could explain bluer emission, was indeed inferred by \cite{beri_black_2019}. Meanwhile, jet base synchrotron emission has previously been shown to primarily emit in the red optical band (\citealt{Fender_Rapid_1997}, \citealt{Gandhi_Variable_2011}). If obscuring material were then to pass in front of the primarily red jet base, very close to the black hole, it may preferentially highlight the comparatively bluer disc and create the change in colour seen during these dips.
	
	
	Our DCFs from Section \ref{ccf} are relevant to this discussion. In Figure \ref{fig:dcf50}, a weak but significant correlation can be seen directly after 0s in every band. Such a feature has been seen in numerous LMXB systems before, and has been linked to synchrotron emission from an inner jet \citep{gandhi_elevation_2017}. The greater strength of this feature in $r'$ compared to $g'$ also supports this interpretation. However, this feature is extended, instead of the sharp peak that is predicted for such a phenomenon. This could point towards X-ray reprocessing, but it is also important to note that most of the literature on this source does not support a reprocessing scenario (\citealt{armas_padilla_multiwavelength_2013}, \citealt{weng_multiwavelength_2015}, \citealt{beri_black_2019}, \citealt{qiao_model_2013}). Also, other sources that are thought to emit synchrotron emission, such as V404 Cyg, have shown distinct, rapid red flares in their lightcurves \citep{gandhi_furiously_2016}; these are not present in the lightcurve of J1357.

	\subsection{Optical Dips -- X-Ray Response} \label{discussdips}
	
	It has been previously shown that there is a lack of any discernible response in X-rays to the optical dips. Our data supports these findings through superposition analysis (Section \ref{dips}) and the X-ray PSD (Section \ref{psd}). A single RXTE QPO was seen in the 2011 outburst at similar frequencies, but it did not strongly match with the observed evolution (JCS13, \citealt{armas_padilla_x-ray_2014}).
	
	To further test if any X-ray dips (correlated with the optical) were present and merely hidden by poor S/N, we created a simulation based on the ULTRACAM $u'$ data. We first took the lightcurve as a baseline, and binned it to correspond to that in X-rays. We then scaled this baseline down in count rate so that its mean matched the X-ray mean rate, and then simulated random sampling around this baseline using Poisson statistics. This was done five hundred times, a CCF was made between the original baseline and each simulation, and the results were then combined.
	
	The resultant CCF was found to show a strong correlation both before and after zero lag at a correlation coefficient of 0.3, far stronger than that seen in Figure \ref{fig:dcf50}. From this, we draw the conclusion that even at low photon count rates, we would have expected to detect dips of similar size to the optical, were they present in the X-ray lightcurve. We therefore conclude that the X-ray lightcurve, for at least the 3 -- 78 keV energy range, appears to be disconnected from the optical in terms of dipping behaviour.

	\subsection{Optical Dips -- Evolution} \label{discussevolution}
	
	JCS13 previously showed how the frequency of the dips declines over the course of the outburst; they theorised that this behaviour was due to the perturbations moving outwards in the disc. Our new data, from both SALT and ULTRACAM (see Figure \ref{fig:dip_evo}), confirm this trend. The original fit given by JCS13, shown in black, is $f$ = 8.91 $\times$ 10$^{-7}$$T^{2}$ -- 12.87 $\times$ 10$^{-4}$$T$ + 46.71 $\times$ 10$^{-2}$, where $T$ is time since the discovery of outburst (in days), and $f$ is measured in Hz. Our new fit, shown in blue and fitted to only the latter four points, is given as $f$ = 8.10 ($\pm$10.39) $\times$ 10$^{-7}$$T^{2}$ -- 2.05 ($\pm$1.24) $\times$ 10$^{-4}$$T$ + 1.37 ($\pm$0.30) $\times$ 10$^{-2}$. The similarity of fits gives further evidence that the dips here are caused by the same phenomenon as in the 2011 outburst. The shift along the x-axis could be explained by the fact that the 2017 outburst was discovered in optical, while the original discovery of the source was made in X-rays.
	
	However, the data that we present shows that the points do not all share the same parabolic pattern that JCS13 presented. This has been hinted at before; \citet{armas_padilla_x-ray_2014} presents an RXTE observation that shows a QPO at 6\,mHz very early on in the outburst (four days after the discovery), before the optical observations by JCS13.
	
	The implication here, combined with our first SALT observation, is that during the early stages in the outburst, the dip frequency \textit{increases} up to a peak before beginning its parabolic decline. With the similarity with our initial SALT observation, we chose not to fit our parabolic trend to that datum.

	\subsection{Spectral Energy Distribution} \label{discusssed}

	\begin{figure}
		\includegraphics[width=\columnwidth]{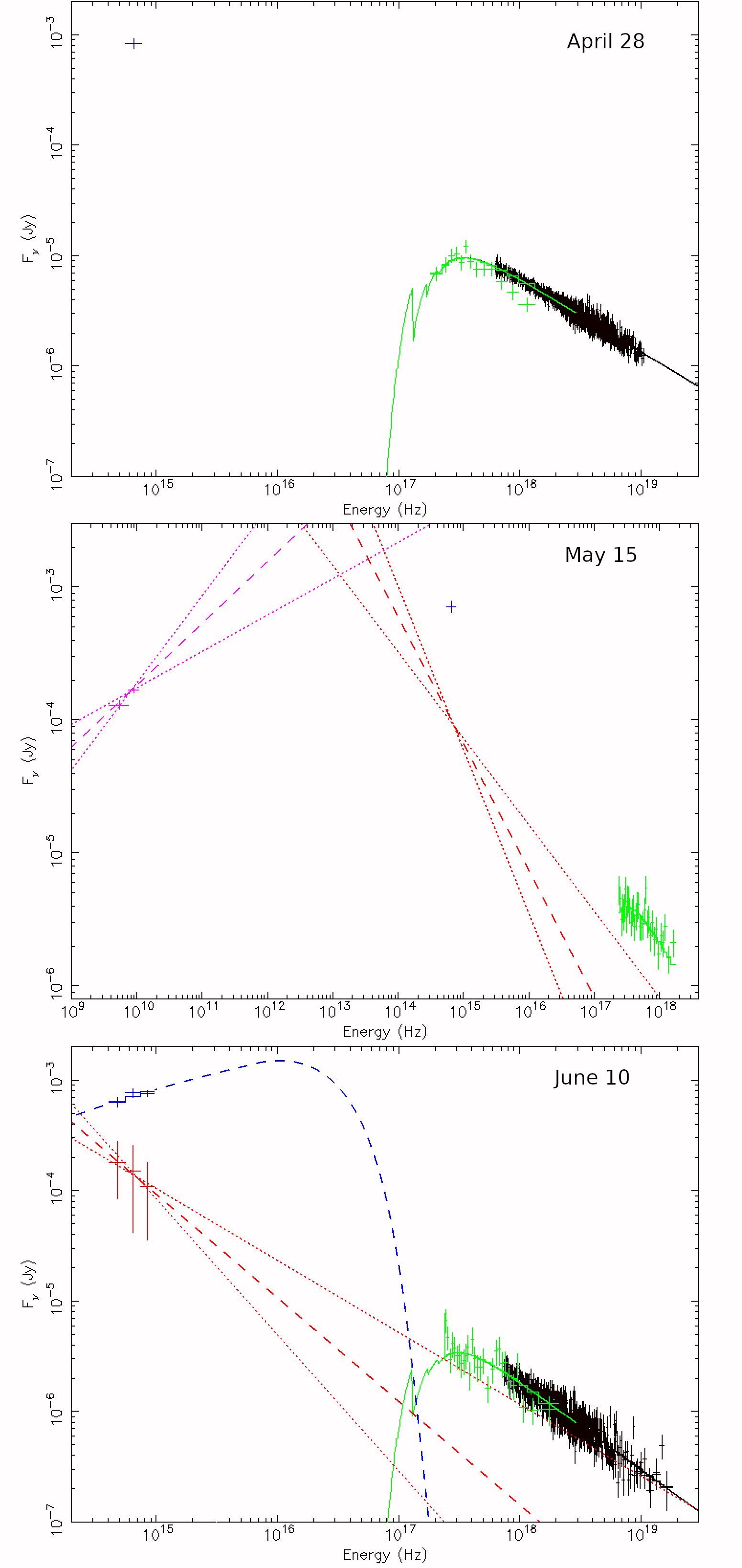}
		\caption{
		SEDs of J1357 from the three nights of correlated observation, showing \textit{NuSTAR} (black), XRT/\textit{Swift} (green), optical (both ULTRACAM and SALT, blue), emission from regions hidden by the dips (ULTRACAM, red),  and ATCA (purple) data.
		{\textbf {Top:}} April 28. X-ray data were fit using a {\tt TBabs} $\times$ {\tt powerlaw} model. The SALT datapoint was not included in this fit.
		{\textbf {Middle:}} May 15. \textit{Swift} data were fit using a {\tt phabs} $\times$ {\tt powerlaw} model, with the photon index frozen at 1.8. ATCA data were fit to a power law (purple dashed line) with errors (dotted lines). The red data were normalised using the SALT datapoint.
		{\textbf {Bottom:}} June 10. X-ray data were fit using a {\tt phabs} $\times$ {\tt powerlaw} model. Disc emission (ULTRACAM data taken during the dips) is shown in blue, and is fitted independently with a {\tt redden} $\times$ {\tt diskbb} model (blue dashed line). Emission from the region hidden by the dips is shown in red, and fitted with a power law (dashed line) with errors (dotted lines).}
		
		\label{fig:SEDComb}
	\end{figure}

	The SEDs shown in Figure \ref{fig:SEDComb} show that the X-ray data can be easily modelled by a pure absorbed power law with no reflection components or dips. This is consistent with a reduced apparent reflection component at high source inclination, and is in accordance with analysis of the June 10 X-ray data carried out by \citet{beri_black_2019}; however, note that \citet{stiele_nustar_2018} did fit a reflection model in their analysis of other X-ray data from this source.
	
	A disk black-body spectrum was used to parameterize the multifilter optical data for this observation, while accounting for intrinsic reddening of the source with the {\tt redden} model; for this latter model, using \citet{Bohlin_Survey_1978}, we set E$_{B-V}$ = 0.0567 mag. The slope of the fit depended both on the optical and X-ray data; since the latter show no sign of disc black body features, we had to ensure that our model did not extend into the X-ray regime. From this, we obtained a range of inner disc temperatures 0.004 $\textless$ T\subscript{in} $\textless$ 0.1 keV. In the SED, a value of T\subscript{in} = 0.05 was plotted as an example.
	
	We also investigated the ULTRACAM broad-band spectral behaviour with respect to the dips. Following methods presented in \cite{hynes_optical_2005}, we fitted the ULTRACAM slope to a power law (of the form $F_{\nu}$ $\propto$ $\nu^{\alpha}$) in log/log space; we did this using data both inside and outside the dips, finding values of $\alpha$ = 0.28 $\pm$ 0.21 (during dips), and $\alpha$ = 0.07 $\pm$ 0.21 (outside of dips). The former is close to the 1/3 value expected for a viscously heated disc - this implies that, during dip events, the source emission is more dominated by the disc. This correlates with previous observations of J1357; \cite{armas_padilla_multiwavelength_2013} found the correlation slope $\beta$ = 0.2 (Between UV/Optical and X-ray) to imply a non-irradiated, viscous accretion disc, while \cite{weng_multiwavelength_2015} stated that a viscously heated disc is the only option for this source, and that the NUV is dominated by emission from the outer non-irradiated viscous disc. We also subtracted the flux observed during the dips from the spectra outside the dips; this difference would be the emission from the region being obscured. Fitting to a power law as before, we found a value of $\alpha$ = -0.94 $\pm$ 0.29; this means that this spectrum is red, which is in line with the source appearing bluer during the dips. This is plotted in the June 10 SED.
	
	With the inclusion of the ATCA data, we can also make inferences to the upper bound of the jet break, by measuring where the radio optically-thick power-law intersects with the optically-thin jet power-law inferred after subtracting the disc from the optical observations. Unfortunately, the ATCA data (May 15) were not simultaneous with the ULTRACAM observations (June 10). However, examining long-term light curves over this period suggests minimal flux changes, at least in the optical and the X-rays. The fluxes that we measure across the three SEDs in \ref{fig:SEDComb} are also approximately constant. We, therefore, make the assumption that there are no substantial changes in terms of jet power-law between May 15 and June 10, and we assume identical jet component fluxes and normalisations between the two dates.

    With this assumption, we measure the synchrotron jet break frequency to be at $\approx$\,1.4\,$\times$\,10$^{13}$ Hz. Accounting for 1-$\sigma$ uncertainties on the slopes, the bounds to the break frequency are 1.97\,$\times$\,10$^{12}$--7.66\,$\times$\,10$^{13}$; these ranges are similar to the jet break frequency of other black hole binaries, e.g. GX 339-4 \citep{Gandhi_Variable_2011} and XTE 1550-564 \citep{Russell_Breaks_2013}.


	\subsection{An Updated Model of J1357} \label{discussmodel}
	
\begin{figure*}
	    \includegraphics[width=\textwidth]{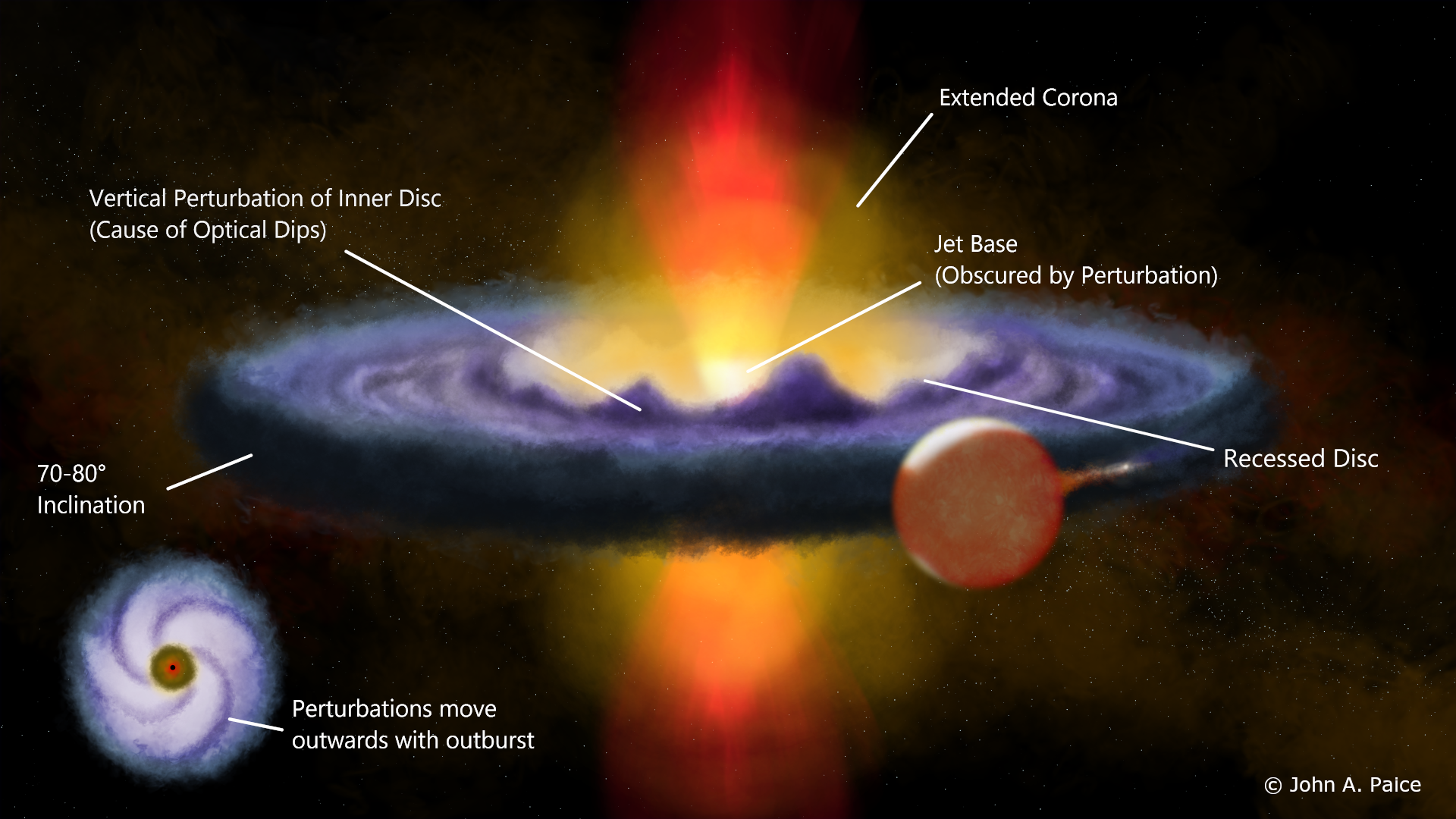}
    	\caption{Schematic of a potential geometry for J1357. We posit that the lack of X-ray dips indicates a large, extended corona, so that any obscuration is minor. Primarily red optical emission from the base of a jet would be obscured by the vertical structure, while a recessed disk would remain mostly unobscured, explaining the colour of the resultant dips. {\textbf {Inset:}} Face-on schematic, showing the outward-moving structure (appearing like spiral arms) in the disc.}
		\label{fig:schematic}
	\end{figure*}

	Based on our new data, we put forward a model for J1357 which could account for the source's observed properties. We illustrate this model in Fig. \ref{fig:schematic}.
	
	We require a multi-component model to explain the spectra and colours. With this in mind, we postulate that this source features a truncated disc, with an inner radius that is recessed from R$_{ISCO}$. Between the disc and the black hole, we suggest that there is an extended X-ray corona, the spherical nature of which making it visible high above any perturbations.
	
	We also propose a region of jet emission near the black hole itself, made up of a superposition of several smaller regions, primarily emitting in red wavelengths. This region, facilitated by a high inclination, would be sporadically occluded by vertical extensions of the accretion disc. When this occurs, less red light is seen, giving preference to the bluer disc. Over the course of the outburst, the perturbations in the accretion disc first move inwards to some minimum distance, and then propagate outwards; with the changing orbit that these perturbations thus have, the frequency of the dips they cause changes accordingly. It is possible that this changing orbit could echo a changing truncation radius of the disk.
	
	The ways that this model differs from the one presented in JCS13 is the addition the truncated disc, the jet base, and the clarification of what the perturbations occlude. There are several advantages to these changes; first, the truncated disc allows for an extended X-ray corona, which helps to explain the apparently low source luminosity, the lack of X-ray reflection features (due to there being no hot inner disc for the X-rays to reflect off of), the lack of any X-ray dips (due to the corona extending far above any perturbations, preventing any occlusion), and the low absorption due to the disc (due to most X-ray emission not passing through the disc), while still having a high source inclination ($\geq$70$^{\circ}$); this solves the inclination issues which were raised primarily by JCS13, \citet{beri_black_2019}, and \citet{torres_vlt_2015}.
	
	Along with the extended corona, the superposition of jet emission regions could explain the optical response to X-ray emission seen in the DCFs; the correlation up to a lag of several seconds could be due to the superposition of jet emission regions, where emission takes place over time as material from the disc travels through successive zones - this would result in a `smearing' of usual features, which would also compromise fast variability (as seen in other LMXB sources). Such a red-emitting jet emission region, when occluded by the vertical extensions mentioned in the previous model, would cause the source to appear bluer - as seen in the dips presented in this paper.
	
	It is worth noting the companion star in our model. Continuing the idea put forward by JCS13, we also assume that the companion star does not extend (at least meaningfully) above the disc rim. This would therefore explain the lack of eclipses that we would otherwise expect to see.
	
	These features, if present, could be confirmed in future observations; for example, a jet emission region could feature a corresponding synchroton self-Compton component in the X-ray. Higher resolution optical and X-ray data, that newer generation instruments would be invaluable in providing, could also see signs of jet emission in more accurate DCFs. Additionally, the presence of jets would be detectable by radio observations of this source during outburst. X-ray dips could also be present in lower-energy lightcurves of the source.
	
	This model also suggests that the source is intrinsically faint in X-rays. If it represents the tip of the iceberg of a larger faint population, it will likely be difficult to identify them; however, surveys like LSST should help \citep{Johnson_LSST_2019}.
	
	Swift J1357 remains complex, and this proposed scenario does not account for every feature. For example, it does not explain why the jet emission region would be extended in such a manner, nor why the inner disk is truncated, nor why the perturbations move outwards during outburst. Another aspect not investigated here is any impact of emission and absorption line contributions. For this, we would need time-resolved optical spectra that resolve the dips. Therefore, further multi-wavelength investigations of this source during outburst are highly desirable, particularly with higher-resolution X-ray and optical timing instruments for epochs on a similar timescale or longer, allowing for more refined DCFs and probing of X-ray variability. These observations are essential for understanding the physical processes responsible for this highly unusual behaviour.

	\section{Conclusions}
	
	We have investigated new ULTRACAM, ATCA, SALT, \textit{Swift} and \textit{NuSTAR} observations from the 2017 outburst of J1357, a number of which were were carried out simultaneously. In every optical observation, we clearly see the optical dips that were reported previously, and once again show no detectable X-ray dipping.
	
	In applying a Fourier analysis and fitting Lorentzians to both ULTRACAM and SALT data, we found the frequency of the dips to evolve over time, matching the decreasing pattern reported by JCS13; in our observations, this frequency changes from $\sim$1 $\times$ 10$^{-2}$ Hz to $\sim$2 $\times$ 10$^{-3}$ Hz (period range 100-500s) over the course of 82 days.
	
	In the ULTRACAM + \textit{NuSTAR} simultaneous observation, these dips were found to have an average V-shaped pattern (reminiscent of eclipsing scenarios) that lasts $\sim$150 seconds. In analysing the colour of these features, we found that longer wavelengths are more affected than shorter ones, giving the source a 'bluer' colour during the dips - this is contrary to what would be expected if they were caused by standard dust obscuration. This relation was clearly seen in data binned every 10 seconds. Thus, we rule out standard dust as the sole cause of features longer than that time period.
	
	Analysis of the source SED reveals that the optical emission cannot be fully explained by reprocessing, implying that a significant part of the optical emission is likely to be intrinsic emission from the disc, in agreement with other results.
	
	We also applied a discrete correlation function between the coincident optical and X-ray lightcurves. While a small peak at short positive lags was found, reminiscent of synchrotron emission from a jet, the main result of the DCF is the lack of a strong lag signature, implying that there is no strong link between the the X-ray and optical variabilities.
	
	Considering this lack of significant correlation, combined with no obscuration in the X-rays and a reconfirmation of the changing timescale of the dips over the outburst, we propose a possible geometry outlined in Figure \ref{fig:schematic}, updated from current models to fit these results. The salient features of this model include a high inclination angle, an extended X-ray corona, a bluer outer recessed disc, and clumpy obscuring regions within that disc occulting a relatively red inner emitter that could be a jet.

	
	\section*{Acknowledgements}
	
	
	We would like to thank the ATCA, NTT, SALT, \textit{Swift} and \textit{NuSTAR} teams for their valuable work, as well as Richard Ashley and David Sahman for carrying out the observations. Thanks go to A. Veledina for their discussion on correlated X-Ray and optical Timing, and to P. Boorman for discussions on this topic and help with the SED. Much thanks also go to C. Done and M. Kimura for their help with SED fitting. Thanks also go to Greg Sivakoff for their help with coordination efforts, as well as SmartNET for its part in coordinating observations. We would also like to thank the referee for their helpful comments and suggestions.
	
	JP would like to thank the University of Southampton and IUCAA for facilitating this project, and is part supported by funding from a University of Southampton Central VC Scholarship. JP also acknowledges support from the UGC-UKIERI Phase 3 Thematic Partnerships.
	
	PAC acknowledges support from the Leverhulme Trust.
	
	VSD and ULTRACAM acknowledge the support of the STFC.
	
	AB is grateful to both the Royal Society and to SERB (Science and Engineering Research Board), India, for financial support through Newton-Bhabha Fund. A.B. is also supported by an INSPIRE Faculty grant (DST/INSPIRE/04/2018/001265) by the Department of Science and Technology, Govt. of India.
	
	DA thanks the support of the Royal Society.
	
	DB acknowledges research support through the National Research Foundation (NRF) of South Africa and would like to thank International Space Science Institute, Beijing, for support under the auspices of an International Team meeting.
	
    JCAM-J is the recipient of an Australian Research Council Future Fellowship (FT140101082).
    
	The Australia Telescope Compact Array is part of the Australia Telescope National Facility which is funded by the Australian Government for operation as a National Facility managed by CSIRO.
	
	The European VLBI Network is a joint facility of independent European, African, Asian, and North American radio astronomy institutes. Scientific results from data presented in this publication are derived from the following EVN project code(s): RM010. This work has made use of data from the European Space Agency (ESA) mission {\it Gaia} (\url{https://www.cosmos.esa.int/gaia}), processed by the {\it Gaia} Data Processing and Analysis Consortium (DPAC, \url{https://www.cosmos.esa.int/web/gaia/dpac/consortium}). Funding for the DPAC has been provided by national institutions, in particular the institutions participating in the {\it Gaia} Multilateral Agreement.
	
	Some of the observations reported in this paper were obtained with the Southern African Large Telescope (SALT) under the Large Science Programme on transients (2016-2-LSP-001). Polish support of this SALT programme is funded by grant no. MNiSW DIR/WK/2016/07.
	
	This research has made use of data, software and/or web tools obtained from the High Energy Astrophysics Science Archive Research Center (HEASARC), a service of the Astrophysics Science Division at NASA/GSFC and of the Smithsonian Astrophysical Observatory's High Energy Astrophysics Division.
	
	This work made use of data supplied by the UK Swift Science Data Centre at the University of Leicester.
	
	This research has made use of the
    \textit{NuSTAR} Data Analysis Software (NuSTARDAS) jointly developed by the ASI Science Data Center (ASDC, Italy) and the California Institute of Technology (Caltech, USA).
	
	This research has also made use of NASA's Astrophysics Data System.
	
	
	
	
	\bibliography{main}
	\bibliographystyle{mnras}
	
	
	
	
	
	
	

	\bsp 
	\label{lastpage}
\end{document}